\documentclass[%
 reprint,
 amsmath,amssymb,
 aps,
 longbibliography,
showkeys
]{revtex4-1}

\usepackage[resetlabels]{multibib}
\newcites{sec}{Reference}

\usepackage{graphicx}% Include figure files
\usepackage{algorithm}
\usepackage{algorithmic}
\usepackage{dcolumn}% Align table columns on decimal point
\usepackage{hyperref}
\hypersetup{colorlinks=true, citecolor=blue, urlcolor=blue, linkcolor=blue}
\usepackage{bm}% bold math
\usepackage{epstopdf}
    \usepackage{titlesec}
       \titleformat{\chapter}[display]
             {\normalfont\Large\bfseries}{\thechapter}{11pt}{\Large}
       \titleformat{\section}
             {\normalfont\large\bfseries}{\thesection}{11pt}{\large}
       \titlespacing*{\chapter}{0pt}{0pt}{15pt} %left, beforesep, aftersep, right
       \titlespacing*{\section}{0pt}{3.5ex plus 1ex minus .2ex}{2.3ex plus .2ex}
\usepackage[titletoc]{appendix}
\usepackage{booktabs}
\usepackage{multirow}

\begin{document}

\preprint{APS/123-QED}

\title{Robust and Efficient Network Reconstruction in Complex System \\ via Adaptive Signal Lasso}

\author{Lei Shi$^{1,2}$}
\email{lshi@ynufe.edu.cn}
\author{Jie Hu$^3$}
\author{Libin Jin$^2$}
\author{Chen Shen$^4$}
\author{Huaiyu Tan$^1$}
\author{Dalei Yu$^1$}
%\email{w-zhen@nwpu.edu.cn}
%\author{Stefano Boccaletti$^{4,5,6}$}

\affiliation{
	1. School of Statistics and Mathematics, Yunnan University of Finance and Economics, Kunming, 650221, China. \\
	2. Interdisciplinary Research Institute of Data Science, Shanghai Lixin University of Accounting and Finance, Shanghai 201209, China.\\
	3. School of Management, University of Science and Technology in China, Hefei 230026, China \\
	4. Faculty of Engineering Sciences, Kyushu University, Kasuga-koen, Kasuga-shi, Fukuoka 816-8580, Japan }
	
	% HU Jie email: hujie@mail.ustc.edu.cn
	
%	3. Center for OPTical IMagery Analysis and Learning (OPTIMAL), Northwestern Polytechnical University, Xi'an 710072, China.\\
%	4. CNR - Institute for Complex Systems, Via Madonna del Piano 10, 50019 Florence, Italy.\\
%	5. Unmanned Systems Research Institute, Northwestern Polytechnical University, Xi'an, 710072, China.\\
%	6. Moscow Institute of Physics and Technology, Institutskiy per., Dolgoprudny, Moscow Region, 141701 Russia.}
\date{\today}

\begin{abstract}

Network reconstruction is   important   to the understanding and control of collective dynamics in complex systems. Most   real networks exhibit sparsely connected properties, and the connection parameter is a signal (0 or 1). Well-known shrinkage methods such as lasso or compressed sensing (CS) to recover structures of complex networks cannot suitably reveal such a property; therefore, the signal lasso method was  proposed recently to solve the network reconstruction problem and was found to outperform lasso and CS methods. However,   signal lasso suffers the problem that the estimated coefficients that fall   between 0 and 1 cannot be successfully selected to the correct class. We propose a new method, adaptive signal lasso, to estimate the signal parameter and uncover the topology of complex networks with a small number of observations. The proposed method has   three advantages: (1) It can effectively uncover the network topology with high accuracy and is capable of completely shrinking the signal parameter to either 0 or 1, which eliminates the unclassified portion in network reconstruction; (2) The method   performs   well   in scenarios of both sparse   and dense signals and is robust to noise contamination; (3) The method only needs to select one tuning parameter versus  two  in signal lasso, which greatly reduces the computational cost and is easy to apply. The theoretical properties of this method are studied, and   numerical simulations from linear regression, evolutionary games, and Kuramoto models are   explored. The method is illustrated with  real-world examples from a human behavioral experiment and a world trade web.

\end{abstract}

\maketitle

\section{Introduction}

Complex networks have wide applications  and seen much progress \cite{strogatz2001exploring,barabasi2012,albert2002statistical,boccaletti2006complex}. In a complex network, the pattern of   node-to-node interaction or   network topology is unknown. To uncover the network topology based on a series of observable quantities obtained from experiments or observations is  important and may play a   role in the understanding and controlling of collective dynamics of complex systems \cite{han2015robust,wang2011network,peixoto2018reconstructing}. Network reconstruction as an inverse problem in network science has been paid much attention recently, such as in the reconstruction of gene networks using expression data    \cite{gardner2003inferring,geier2007}, extraction of various functional networks in the human brain from activation data    \cite{grun2002,supekar2008network}, and detection of organizational networks in social science and trade networks in economics \cite{squartini2018}. Evolutionary game-based dynamics   have been used to study   network reconstruction, where it is possible to observe a series of a small number of discrete   quantities  \cite{wang2011network,han2015robust,shi2020recovering,shi2021}, in which case the problem can be transformed to a statistical linear model with sparse and high-dimensional properties.

We use two typical  examples to illustrate how such signal parameters appear in  practice.
The first example is a dynamic equation governing the evolution state in a general complex system, which can be written as   differential equations \cite{wang_wx2016,raimondo2021measuring}
\begin{equation}\label{eq0}
\dot{\bf x}_i(t)=\psi_{i}({\bf x}_i(t),\nu_i)+\sum_{i=1}^N a_{ij}\phi_{ij}({\bf x}_i(t),{\bf x}_j(t))+\epsilon_i(t),
\end{equation}
where ${\bf x}_i(t)$ denotes an $m$-dimensional internal state variable of a system consisting of $N$ dynamic units at time $t$, where $\psi_{i}\in R^m$
and $\phi_{ij}$  $\in R^m$, respectively, define the intrinsic and interaction dynamics of the units; $\epsilon_i(t)$ is a dynamic noise term; 
$\nu_i$ is a set of dynamic parameters; and $a_{ij}$ defines the interaction topology and is called by an adjacency matrix such that $a_{ij}$ = 1 if there is a direct physical
interaction from unit $j$ to $i$, and $a_{ij}$ = 0 otherwise. The matrix $A=[a_{ij}]$ completely defines a network with size $N$,  i.e., an abstraction used to model a system that contains discrete interconnected elements. The elements are represented by nodes (also called vertices), and connections   by edges. In general, ${\bf x}_i(t)$ can be observed as time-series data, but $a_{ij}$ for $i=1, \cdots, N$ are unknown and must be estimated. It is clear that Eq. (\ref{eq0}) can be rewritten as a linear regression model if the functional forms of $\psi_{i}$
and $\phi_{ij}$ are known. This model includes   synchronization models, oscillator networks, and spreading networks \cite{wang_wx2016}.

The second example comes from the evolutionary game on structured populations, where a node represents a player, and a link indicates that two players have a game relationship. The prisoner's dilemma game (PDG), snowdrift game (SDG), or spatial ultimatum game (SUG) can be used for   network reconstruction \cite{wang2011network,han2015robust,peixoto2018reconstructing,shi2021}. We use the PDG, with   temptation to defect $T$, reward for mutual cooperation $R$, punishment for mutual defection $P$, and sucker's payoff $S$. Thus, the payoff matrices can be defined as
\begin{equation}\label{eq1}
\mathbf{M_{PDG}} =
\left( \begin{array}{cc}
R \ \ S \\
T \ \ P
\end{array}\right),
\end{equation}
and   $T>R>S>P$, where mutual defections  are the equilibrium solutions  \cite{santos2005scale,nowak1992evolutionary,hauert2004spatial,szabo2007evolutionary}. For simplicity, these parameters are re-scaled as $T=b$, $P=S=0$, and $R=1$, where $1 \leq b < 2$ ensures a proper payoff ranking ($T>R>P>S$) and captures the essential social dilemma \cite{nowak1992evolutionary,perc2017statistical}.  In some experimental driving studies, each player can interact with other players by choosing either a cooperator (C) or defector (D) to obtain their payoff and the procedure is continued for a  predetermined number  of rounds\cite{li2018punishment,wang2017onymity,wang2018exploiting,shi2020freedom}. In a theoretical study, some updating mechanism can be used to generate theoretical data on three types of topologies \cite{szabo2005phase}.  Suppose that each player $i$ is either   a cooperator (C) or defector (D) with equal probability, which can be written as  $s_i=(1,0)$ or $s_i=(0,1)$. In a spatial PDG game, player $i$, say the focal player, acquires its fitness (total payoff) $F_i$ by playing the game with all its connected neighbors,
\begin{equation}\label{eq3}
F_i=\sum_{j \in \Omega_i} s_{i}M_{PDG}s'_{j}=\sum_{j=1, j \neq i}^N a_{ij} P_{ij},
\end{equation}
where $\Omega_i$ is the set of all connected neighbors of player $i$, $P_{ij}=s_{i}M_{PDG}s'_{j}$.  
Eq. (\ref{eq3}) can be converted to a linear model, with elements $a_{ij}$   of the adjacency matrix for a   network. If $a_{ij}=1$, then  players $i$ and $j$ are connected, and if $a_{ij}=0$, then they are not. The process can   produce time-series data. In each step, players can update their strategies using a rule, or determine them themselves.  Suppose $L$ accessible time is available. Then, the model   containing the time-series data can be rewritten as
\begin{equation}\label{eq5}
Y_i={\Phi}_i \tilde X_i+e_i,
\end{equation}
where $Y_i=(F_i(t_1), F_i(t_2), \cdots, F_i(t_L))' $, ${\Phi}_i=[P_{ij}(t)]\in R^{L\times (N-1)}$,
and $\tilde X_i=(a_{i1}, \cdots, a_{i,i-1}, a_{i,i+1}, \cdots, a_{iN})'$, in which the $i$th connection with itself is removed. The introduction of $e_i$ is due to noises or missing nodes in real applications. Therefore the aim of Eq.  (\ref{eq5}) is to estimate the elements $a_{ij}$ of the connectivity matrix, which is important for uncovering network structures, such as possible  social network in the social science or an intrinsic scientific relationship in gene-regulatory network reconstruction from the expression data in systems biology. 

In  the areas of complex systems and applied physics, the compressed sensing (CS) or lasso methods are   techniques to estimate $a_{ij}$ and achieve the purpose of network reconstruction  \cite{wang2011network,han2015robust,shi2020recovering}, and  lasso has been found to be robust against   noises  in the reconstruction of sparse signals.    Player $i$ and player $j$ are predicted to have a game   relationship (connection)  if $\vert \hat a_{ij} -1\vert \leq 0.1$, and no   relationship if $\vert \hat a_{ij}\vert \leq 0.1$, where $\hat a_{ij}$ is an estimator of $a_{ij}$. Otherwise, the   relationship is not identifiable.  Although the CS or lasso method can shrink   parameter estimates toward   zero under   natural sparsity in complex networks,  links between nodes cannot be shrunk to a true value of 1, which will   decrease   estimation accuracy in most cases. For this reason, Shi {\it et al.} (2021) \cite{shi2021} proposed the signal lasso method to solve the network reconstruction problem and   found that it performed better than the lasso and CS methods. However   values of $\hat a_{ij}$ that fall in the interval (0.1,0.9) cannot be placed  in  the correct class and leave an unclassified portion in network reconstruction.

We propose a   method called adaptive signal lasso to estimate the signal parameter and uncover the   topology in complex networks with a small number of observations. The idea  is to add  weights to the penalty terms of signal lasso. We show that our method can shrink the parameter to either 0 or 1 completely and greatly improve estimation accuracy. Furthermore, this   method only tunes one parameter in a very small range, which saves computation time. The method is   robust against noise and missing nodes, since a least square error control term is included.  We conduct   simulations and comparison studies through a linear regression model with signal parameters using six current shrinkage methods and our proposed method. We   validate our reconstruction framework through an evolutionary game and synchronization model by considering three topological structures: random (ER), small word, and scale-free. The results show that our method can achieve  high prediction accuracy compared with the other methods, remove the unclassified portion of subjects, and decrease the computational cost. We use two real examples for illustration and find that our method  performs surprisingly well at detecting signals in all cases (including case of dense networks). Our   method has potential applications in  fields such as  social, economic, physical, and biological systems, and the recovery of hidden networks.

\section{Motivation}

Consider the general linear regression model
\begin{equation}\label{eqa01}
Y=\Phi X+\epsilon
\end{equation}
where $\epsilon$ is a noise or random error with mean zero and finite variance, $\Phi = [\phi_{ij}]$ is an $n\times p$ matrix, $Y = [y_i]$
is an $n\times 1$ vector, and $X = [X_i]$ is a $p \times 1$ unknown vector. To eliminate the intercept from (\ref{eqa01}), throughout this paper, we center the response   and predictor variables so that the mean of the response is zero. We assume the parameter $X$ has a signal property, e.g., the true values of $X_j$, $j=1,\cdots, p$, are either 0 or 1. This kind of problem is   common in the reconstruction of complex networks to identify a signal as either connected  or not   \cite{han2015robust,wang2011network}.

The signal lasso method    minimizes  \cite{shi2021} 
\begin{equation}\label{eqa03}
\frac{1}{2}\sum_{i=1}^n (y_i-\sum_{j=1}^p \phi_{ij}X_j)^2+\lambda_1\sum_{j=1}^p\vert X_j\vert +\lambda_2\sum_{j=1}^p\vert X_j -1\vert,
\end{equation}
where  $\lambda_1, \lambda_2 > 0$  are two tuning parameters. 
The  term   $\sum_{j=1}^p\vert X_j -1\vert$ is added in the penalty term because some elements of $X$ should be 1. When $\lambda_2=0$, it reduces to the lasso method  \cite{tibshirani1996regression}. The tuning parameters $\lambda_1$ and $\lambda_2 > 0$ must be determined by the  dataset  through cross-validation.
This is a compromise between shrinking terms  to 0 and 1, and we expect some elements of $X$ will be close to 0, and others to 1. 

If the columns of $\Phi$ are orthogonal to each other and $p < n$,
denoting the ordinary least squares estimate   by $\hat X_{0}=\Phi'Y$,
the estimator of $X$ in signal lasso is
\begin{equation}\label{eqa08}
\hat X_k=\left\{
\begin{array}{l}
(\hat X_{k0}+\delta_1)_{-}, \ \ \ \ \ \ \ \hat X_{k0}\leq0, \\
(\hat X_{k0}-\delta_2)_{+}, \ \ \ \ \ \ \ 0<\hat X_{k0}\leq 1+\delta_2, \\
\max\{1, \hat X_{k0}-\delta_1\}, \ \ \hat X_{k0}> 1+\delta_2,
\end{array}\right.
\end{equation}
for $k=1,\cdots, p$, where $\delta_1=\lambda_1+\lambda_2$ and  $\delta_2=\lambda_1-\lambda_2$, and $\hat X_{k0}$ and $\hat X_k$ are the $k$th element of $\hat X_{0}$ and $\hat X$, respectively; $B_{+}$ denote the positive part of $B$, it means that $B_{+}=B$ if $B\ge 0$ and 0 otherwise. $B_{-}$ is similarly defined as the negative part of $B$.

\begin{figure}
\centering{\includegraphics[width=0.45\textwidth]{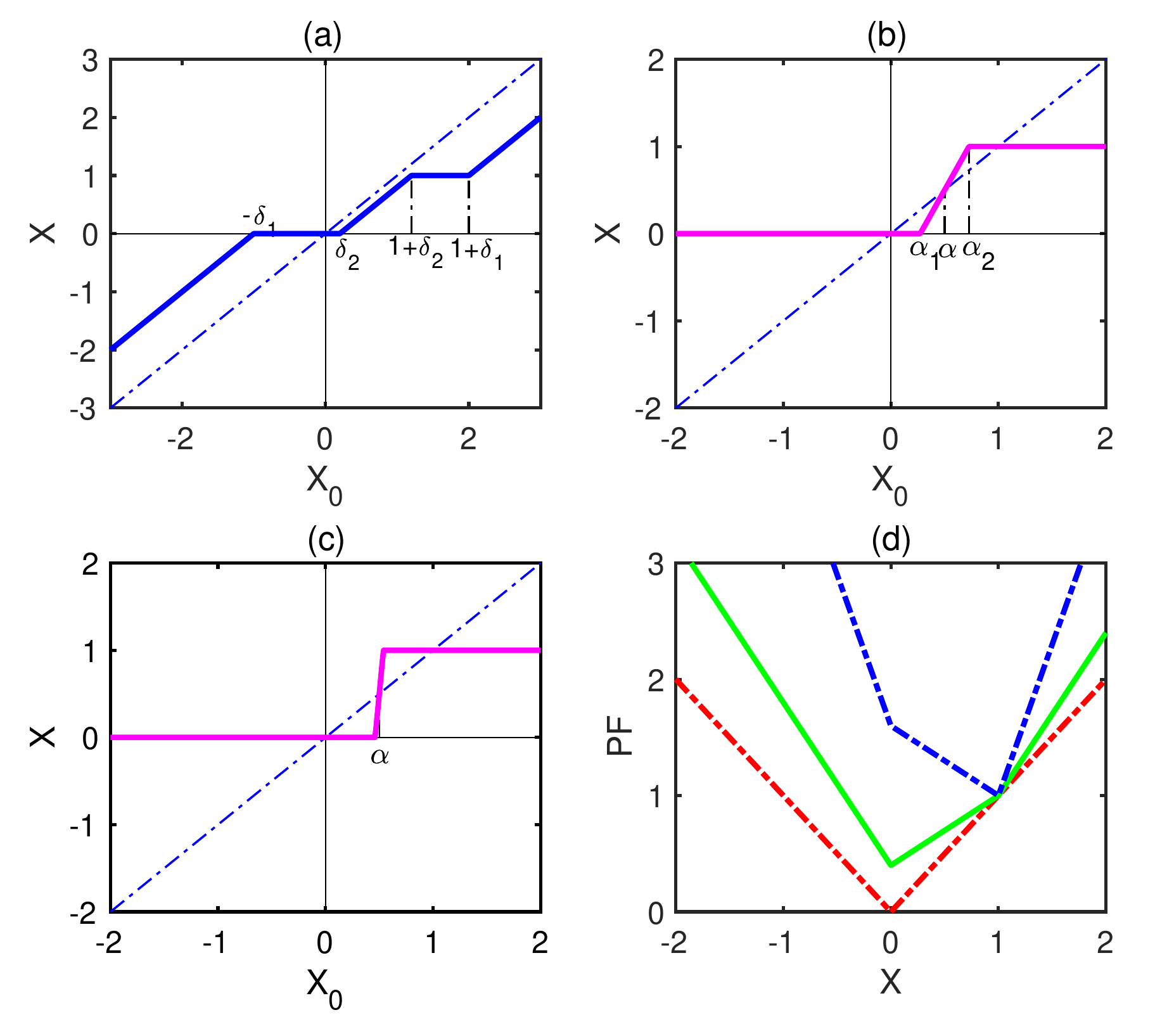}}
\caption{Solution of $X$ under an orthogonal design in signal lasso and adaptive signal lasso.  (a) signal lasso with $\lambda_1=0.6$ and $\lambda_2=0.4$, where $\delta_1=\lambda_1+\lambda_2$, $\delta_2=\lambda_1-\lambda_2$. (b) adaptive signal lasso with $\lambda_1=0.6$ and $\lambda_2=1.2$, where $\alpha=\lambda_1/\lambda_2$, $\alpha_1=\lambda_1/(1+\lambda_2)$, $\alpha_2=(1+\lambda_1)/(1+\lambda_2)$. (d) adaptive signal lasso with $\lambda_1=6$ and $\lambda_2=12$. (d) penalty function of adaptive signal lasso versus $X$ for different OLS of $X$ when $\lambda_1=0.6$ and $\lambda_2=1.2$, where blue line for $X_{10}=0.1, X_{20}=0.9$, green line for $X_{10}=0.9, X_{20}=0.1$, and red line is penalty function of lasso, .}
\label{fig:1}       % Give a unique label
\end{figure}

Fig. \ref{fig:1} (a) shows the solutions of $\hat X$ as a function of $\hat X_0$ under an orthogonal design matrix for the signal lasso method,  where $\lambda_1=0.6$ and $\lambda_2=0.4$, and the $45^\circ$ line of $\hat X=\hat X_0$ is for reference. The signal lasso method not only shrinks the small values of the parameter to zero but also shrinks large values  to 1, and therefore  outperforms the lasso and CS methods in network reconstruction for signal parameters \cite{shi2021}. However, this method still has some unsatisfactory aspects in shrinking signal parameters. First, although   larger values such that $1+\delta_2\le\hat X_{k0}\le 1+\delta_1$ can be shrunk to 1, and   values in the interval $-\delta_1\le\hat X_{k0}\le \delta_2$   to 0,   values in the interval $(\delta_2,1+\delta_2)$   only shift by a constant $\delta_2$, making some parameters unidentifiable. 

 Compared to the pattern shown in Fig. 1a, the pattern shown in Fig. \ref{fig:1}(b) and (c), obtained from our new method, is more favorable, as the middle part between 0 and 1 can be shrunk toward to two directions.    Second, signal lasso involves two tuning parameters, making the computation  costly even if cross-validation   is available. To overcome these   weaknesses,  we propose an efficient modification by given a weight in penalized terms of signal lasso. We find that the estimation of parameters in model in (\ref{eqa01}) can be completely shrunk to 0 or 1, and we
only need to select one tuning parameter, which makes the computation fast and greatly improves its accuracy. 

\begin{figure}
\begin{center}
\centering{\includegraphics[width=0.45\textwidth]{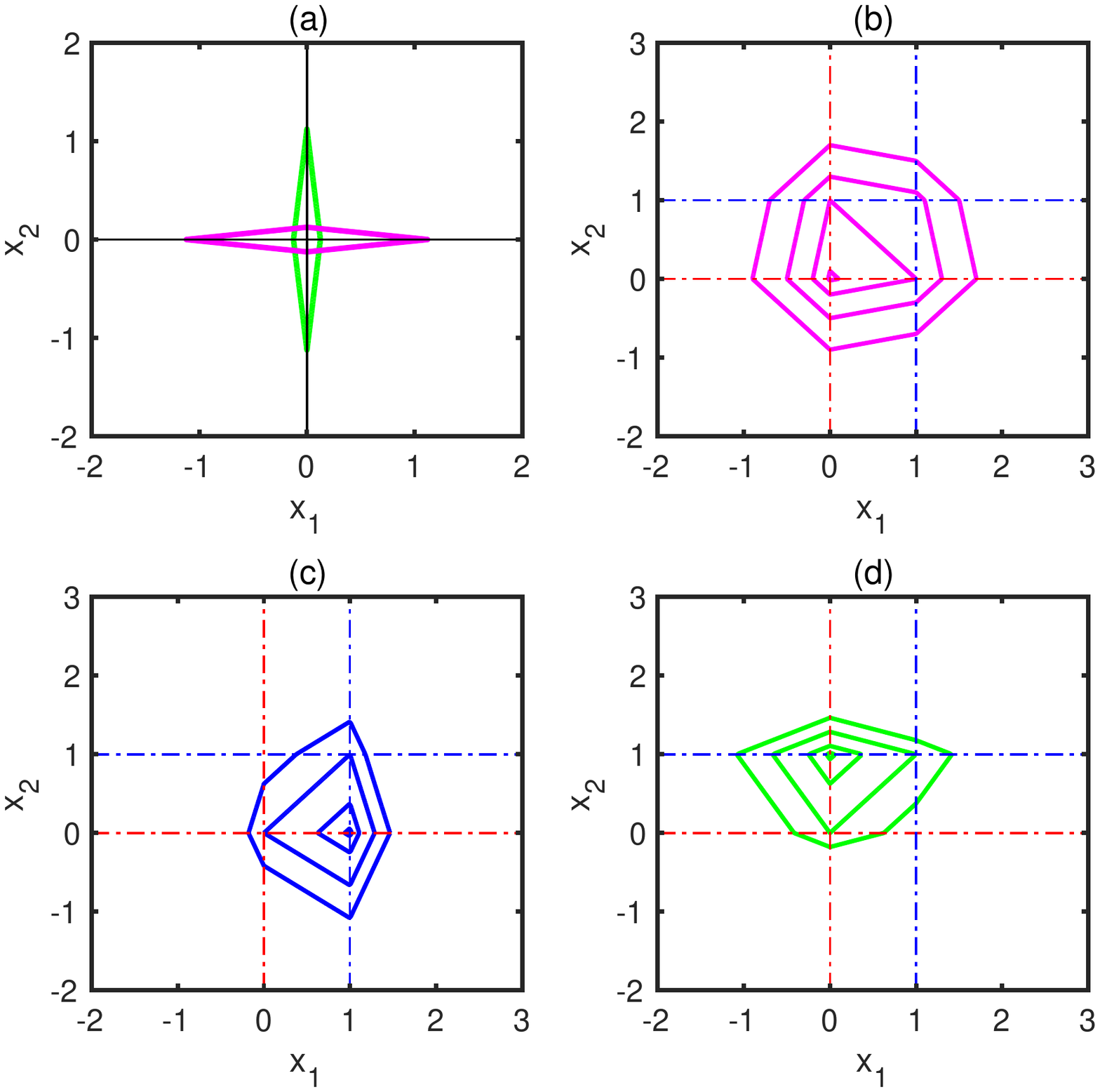}}
\end{center}
\caption{Constraint regions of $PF(x)=c$ for some constant c under four estimation methods in two-dimensional case ($p=2$). (a) Adaptive lasso estimation with penalty function $PF(x)=\sum_{j=1}^2 \vert \hat X_{j0}\vert^{-1} \vert x_j \vert $, c=1, where green line for case of $\hat X_0=(0.1,0.9)$ and magenta line for case of $\hat X_0=(0.9,0.1)$. (b) Signal lasso with penalty function $PF(x)=\lambda_1 \sum_{j=1}^2 \vert x_j\vert+\lambda_2 \sum_{j=1}^2 \vert x_j-1\vert$, $\lambda_1=0.6,\lambda_2=0.4, c=1.7,1.3,1,0.82. $ (c) Adaptive signal lasso with penalty function $PF(x)=\lambda_1 \sum_{j=1}^2 \vert x_j\vert+\lambda_2 \sum_{j=1}^2 \vert X_{j0}\vert \vert x_j-1\vert$, $\lambda_1=0.6,\lambda_2=1.2, \hat X_0=(0.9,0.1),c=1, 0.8, 0.6, 0.5$. (d) Adaptive signal lasso with penalty function $PF(x)=\lambda_1 \sum_{j=1}^2 \vert x_j\vert+\lambda_2 \sum_{j=1}^2 \vert X_{j0}\vert \vert x_j-1\vert$, $\lambda_1=0.4,\lambda_2=0.6, \hat X_0=(0.1,0.9),c=1,0.8,0.6,0.5$.
}
\label{fig:2}
\end{figure}

\section{Adaptive signal lasso}

\subsection{Method}

To deal with the abovementioned problems, we propose following penalized least square function
\begin{equation}\label{eqa12}
L(X\vert\lambda_1,\lambda_2)=\frac{1}{2}||Y-\Phi X||_2^2
+PF(X,\lambda_1,\lambda_2),
\end{equation}
with the penalty function $PF(X,\lambda_1,\lambda_2)$ given by
$$PF(X,\lambda_1,\lambda_2)=\lambda_1 \sum_{j=1}^p \omega_{1j}\vert X_j\vert+\lambda_2 \sum_{j=1}^p \omega_{2j}\vert X_j-1\vert$$
where weight coefficients $\omega_{1j}$ and $\omega_{2j}$  are functions of $\hat X_{j0}$, an initial estimator of $X_{j}$, which can be an ordinary least square estimator for $p<n$, or a ridge estimator for $p>n$. 
A new estimator of $X$ defined by 
\begin{equation}\label{eqa13}
\hat X=\arg\min_{X} L(X \vert \lambda_1,\lambda_2), 
\end{equation} 
 is called  adaptive signal lasso.   For the choice of weights, we expect that the first term of the penalty will have a lower weight, and the second term will have a large weight when $\hat X_{k0}$ is close to $1$ (similar to when $\hat X_{k0}$ is close to $0$).  Motivated by  adaptive lasso, we can choose
that  
$$\omega_{1k}=\vert\hat X_{k0}\vert^{-\nu}, \ \ \ \omega_{2k}=\vert\hat X_{k0}\vert^\gamma$$ 
with $\nu, \gamma>0$ for $k=1,\cdots, p$.  After comparison and analysis, we find the best candidates for the   weights are  $\omega_{1k}=1$ and $\omega_{2k}=\vert\hat X_{k0}\vert$. This is effective in the analysis and will be used throughout this paper if not specified otherwise (see Appendix 1 for  discussions). 

Now, we give the geometry of adaptive signal lasso in the case of an orthogonal design matrix $\Phi$ with $p<n$. After some calculations (see Appendix 2), the   solution is given by
\begin{equation}\label{eqa15}
\hat X_k=\left\{
\begin{array}{l}
\{(1-\lambda_2)\hat X_{k0}+\lambda_1\}_{-}, \ \ \ \ \ \ \ \hat X_{k0}\leq0, \\
\{(1+\lambda_2)\hat X_{k0}-\lambda_1\}_{+}, \ \ \ \ \ \ \ 0<\hat X_{k0}\leq \alpha_2, \\
\max\{1, (1-\lambda_2)\hat X_{k0}-\lambda_1\}, \ \ \hat X_{k0}> \alpha_2,
\end{array}
\right.
\end{equation}
where  $\alpha_2=(1+\lambda_1)/(1+\lambda_2)$. 
In this case,   adaptive signal lasso with $0<\lambda_1<\lambda_2<1$ will enjoy satisfactory technical properties.
Fig. \ref{fig:1}(b) shows the solutions of $\hat X$ as a function of $\hat X_0$ for adaptive signal lasso in the special case of (\ref{eqa15}), where the main difference from signal lasso   occurs in the interval (0,1).  It is of interest to see that the values of $\hat X_0$ in $( \alpha_1, \alpha)$ will be shrunk toward   0, while the values of $\hat X_0 \in$ $(\alpha,  \alpha_2)$ will be shrunk toward   1, where $\alpha=\lambda_1/\lambda_2$ and  $\alpha_1=\lambda_1/(1+\lambda_2)$. The line $X=(1+\lambda_2)X_0-\lambda_1$ has a slope of $1+\lambda_2$, which indicates a kind of shrinkage strength. When    $\lambda_2$ increases to 12 and we keep $\alpha=0.5$ (which means $\lambda_1=6$), the pattern given in Fig. \ref{fig:1}(c) shows that almost all parameter estimation in the middle part can be shrunk almost   to 0 or 1. 

Thus we re-parameterize $\lambda_1$ and $\lambda_2$ by $\lambda=\lambda_2$ and $\alpha=\lambda_1/\lambda_2$ and rewrite the the penalty function by
\begin{equation}\label{eqa16}
PF(X,\lambda,\alpha)=\lambda \{ \alpha \sum_{j=1}^p \vert X_j\vert+ \sum_{j=1}^p \vert \hat X_{j0} \vert \vert X_j-1\vert \}.
\end{equation}
It can be easily proved from Eq (\ref{eqa15}) that when $\alpha$  is fixed and let $ \lambda \to +\infty$, then we have
\begin{equation}\label{eqa17}
\hat X_k \to \left\{
\begin{array}{l}
 1, \ \ \ \ \ \ \ \hat X_{k0} > \alpha , \\
 0, \ \ \ \ \ \ \ \hat X_{k0} < \alpha, 
\end{array}
\right.
\end{equation}
since $\alpha_2 \to \alpha$ and $\alpha_1\to \alpha$ in these scenarios. The parameter can be assigned randomly as 0 or 1 if $\hat X_{k0} =\alpha$. This result indicates that if we set $\lambda$ large enough, the estimators from adaptive signal lasso can be completely shrunk to either 0 or 1, thus removing the unidentified set that will be presented in the signal lasso method. Another advantage   is that we only need to select tuning parameter $\alpha$, which dramatically reduces the computation time comparing with signal lasso. In addition, the range for selecting $\alpha$ can be set in a small interval such as $(0.2,0.8)$, since a smaller or larger $\alpha$ is inappropriate in practice. If we do not like   tuning the parameter, a rough value of $\alpha=0.5$ is preferable in most cases in adaptive signal lasso. 
Fig. \ref{fig:1} (d) shows the functional form of the penalty  to show how the penalties behave for different values of  $\hat X_0$. For example, when $\hat X_{10}=0.1, \hat X_{20}=0.9$, the adaptive penalty function tends to shrink toward   (0,1), while it shrinks toward   (1,0) when  $\hat X_{10}=0.9, \hat X_{20}=0.1$. 

%To show some geometries of different lasso methods, Fig.\ref{fig:2} gives constraint regions $PF(X)=t$ for different shrink estimation methods.
%Fig.\ref{fig:2}  (a) lists the graphs for adaptive lasso methods. Fig.\ref{fig:2} (b) is for signal lasso. Fig.\ref{fig:1}  (c)-(d) are for adaptive signal lasso with different values of $(\hat X_{10},\hat X_{20})$  respectively. For example for $\hat X_0=(0.9,0.1)$, the contours for adaptive signal lasso will be centered at point $X=(1,0)$ and gradually converge to this point when $t$ become small. It is of interest to see that shape of  constraint regions $PF(X)=t$ in adaptive signal lasso varies for different target point.

 Fig. \ref{fig:2} shows constraint regions $PF(X)=t$ for different shrink estimation methods.
Fig. \ref{fig:2}  (a) shows the graphs for adaptive lasso methods. Fig. \ref{fig:2} (b) shows signal lasso and Fig. \ref{fig:1}  (c)--(d) show adaptive signal lasso with different values of $(\hat X_{10},\hat X_{20})$ . For example, for $\hat X_0=(0.9,0.1)$, the contours for adaptive signal lasso will be centered at   $X=(1,0)$,  gradually converging to it  when $t$ becomes small. It is of interest to see that the shape of  constraint regions $PF(X)=t$ in adaptive signal lasso varies by target point.

 \subsection{ Algorithm and computation}

It is noted that the penalty function  in Eq (\ref{eqa12}) is  convex. Hence, the optimization problem in (\ref{eqa13}) does not suffer from   multiple local minima, and its global minimizer can be efficiently solved.  We provide an algorithm using the coordinate descent method  \cite{hastie2015book}, an iterative algorithm that updates the estimator by choosing a single coordinate to update, and then performing a univariate minimization over it.
Since $\omega_{1k}$ and $\omega_{2k}$ are known, we denote $\lambda_{1k}^*=\omega_{1k}\lambda_1$, $\lambda_{2k}^*=\omega_{2k}\lambda_2$.
Define a threshold function by
\begin{equation}\label{eqa 20}
 S_{\theta_1,\theta_2}(z)=\left\{
\begin{array}{l}
(z+\theta_1)_{-},  \qquad  z \leq 0,\\
(z-\theta_2)_{+}, \qquad  0< z \leq 1+\theta_2,\\
\max\{1, z-\theta_1 \}, \qquad  z > 1+\theta_2.
\end{array}\right.
\end{equation}
Then, the update can  proceed as
\begin{equation}\label{eqa 23}
\hat X_k^{t+1}\leftarrow S_{\delta_{1k}^*, \delta_{2k}^*}\left(\hat X_k^t+\displaystyle\frac{\left< \hat r^t, \phi_k\right>}{\left< \phi_k, \phi_k\right>}\right)
\end{equation}
where $\delta_{1k}^*=(\lambda_{1k}^*+\lambda_{2k}^*)/\left< \phi_k, \phi_k\right>$, $\delta_{2k}^*=(\lambda_{1k}^*-\lambda_{2k}^*)/\left< \phi_k, \phi_k\right>$, $\left< z_1, z_2\right>$ denotes the inner product of vectors $z_1$ and $z_2$,  $\phi_k$ is the $k$th column of $\Phi$, $\hat X_k^{t}$ is the estimator of $X_k$ in the $t$th step, and $\hat r^t=Y-\Phi \hat X^t$. The   algorithm   applies this update repeatedly in a cyclical manner, updating the coordinates of $\hat X$ along the way. Once an initial estimator of $X$ is given, for example by lasso   or ridge estimation,   updating can   continue until convergence. These results are proved in Appendix 2.

If  Eq. (\ref{eqa13}) is formulated by parameters $\lambda$ and $\alpha$ and $\omega_{1k}=1$, $\omega_{2k}=\vert \hat X_{0k}\vert$, then
$$\delta_{1k}^*=\frac{\lambda(\alpha+\vert\hat X_{k0}\vert)}{\left< \phi_k, \phi_k\right>}, \delta_{2k}^*=\frac{\lambda (\alpha-\vert\hat X_{k0}\vert)}{\left< \phi_k, \phi_k\right>}. $$
Let $\lambda\to +\infty$. It is clear that Eq. (\ref{eqa 23}) will shrink the negative solution to 0 ,and a solution   larger than 1 to 1, because   $\delta_{1k}^* \to +\infty$. For $\hat X_{k0}\in (0,1)$, when $\hat X_{k0}>\alpha$ and $\lambda\to +\infty$, then $\delta_{2k}^* \to -\infty$ and only the last condition in Eq. (\ref{eqa 20}) holds. Therefore, the solution gives a result of 1, since $\delta_{1k}^* \to +\infty$ in this case. When $\hat X_{k0}<\alpha$ and $\lambda\to +\infty$, then $\delta_{2k}^* \to +\infty$, and only the first two conditions in Eq.(\ref{eqa 20}) are possible. However, they both equal   0 since $\delta_{1k}^* \to +\infty$.  This indicates that the conclusion given in Eq. (\ref{eqa15})   still holds  in the general case, which will be helpful for selecting the tuning parameters in adaptive signal lasso.

\subsection{Tuning the parameter}\label{sec:ctp}

  From the previously mentioned properties based on $\lambda$ and $\alpha$ in Eq.(\ref{eqa16}), we can specify a large value for $\lambda$ and only tune $\alpha$ using cross-validation (CV), which only involves one parameter and will greatly reduce   computation. Furthermore, since $\alpha$ represents  the proportion of data compressed to 0 in the interval (0, 1), it should be less than 1 and greater than 0. In practical situations,  too small or large an $\alpha$ is not preferable; hence an empirical treatment is to select $\alpha$ in a small interval such as (0.2,0.8). Our simulation confirms this is enough to conduct cross-validation. In addition, if one does not like to tune the parameter in adaptive signal lasso, an approximate value of $\alpha=0.5$ can be used. We use $\lambda=1000$ as a large value, and find it is good enough in our calculations, and remove the unclassified portion in  network reconstruction. 
   
\begin{table*}
\caption{Measures for accuracy of network reconstruction }
\label{tab: illus}
\begin{center}
\begin{tabular*}{400pt}{@{\extracolsep\fill}cccccccccccc@{\extracolsep\fill}} 
\toprule
\hline
\multicolumn{2}{c}{\multirow{2}{*}{Actual class}}&\multicolumn{9}{@{}c@{}}{Predicted class } \\ 
\cline{3-11} 
& &\multicolumn{3}{@{}c@{}}{Signal }&\multicolumn{3}{@{}c@{}}{Non signal }&\multicolumn{3}{@{}c@{}}{Unclassified}\\  \hline
&Signal class &   \multicolumn{3}{@{}c@{}}{True positive (TP) }&\multicolumn{3}{@{}c@{}}{False negative (FN) }&\multicolumn{3}{@{}c@{}}{Unclassified positive (UCP)} \\
&Non signal class &   \multicolumn{3}{@{}c@{}}{False positive (FP) }&\multicolumn{3}{@{}c@{}}{True negative (TN) }&\multicolumn{3}{@{}c@{}}{Unclassified negative (UCN)} \\
\hline
\bottomrule
\end{tabular*}
\end{center}
\end{table*}

\subsection{The metrics of reconstruction accuracy}\label{sec:3d}

To measure the accuracy of the estimation method, we have to define some metrics in the signal identification problem. As shown in Table \ref{tab: illus}, we adopt  common  notation as in binary classification, where true positive (TP) is the number of   correctly identified true signals, true negative (TN) is the number of correctly identified  non-signals, false positive (FP) is the number  of non-signals incorrectly identified as  signals, and false negative  (FN) is the number of signals incorrectly identified as  non-signals. However, in our analysis, some lasso-type methods have points that cannot be classified (e.g., the parameter $X$ is classified as  signal if $\hat X\in 1\pm 0.1$, and non-signal if $\hat X\in 0\pm 0.1$, and the remainder are unclassified \cite{wang2011network,han2015robust,shi2021}); hence, we have the additional   classes of unclassified positive (UCP) and unclassified negative (UCN), as shown in Table \ref{tab: illus}.
In traditional classification problems, the predicted class are completely classified to two classes, therefore most common indexes for measuring accuracies are the true positive
rate (TPR, sensitivity or recall), true negative rate (TNR, or specificity) and precision (Positive prediction value, PPV ),  as well as the area under the receiver operating characteristic
curve (AUROC) and the area under the precision recall curve (AUPR)\cite{marbach2012,han2015robust,squartini2018}, where TPR and TNR are defined by 
\begin{equation}\footnotesize\label{eqa 24}
TPR=\frac{TP}{TP+FN}, \ \ \ \  TNR=\frac{TN}{TN+FP}.
\end{equation}
However, these measures have a problem when  size of signals and non-signals are unbalanced  \cite{chicco2017}.  An alternative   solution   employs the Matthews correlation coefficient, 
\begin{equation}\footnotesize\label{eqa 25}
MCC=\frac{TP\times TN-FP\times FN}{\sqrt{(TP+FP)(TP+FN)(TN+FP)(TN+FN)}}
\end{equation}
which correctly takes into account the size of the confusion matrix elements  \cite{chicco2017,chicco2020}. MCC is widely used in machine learning to measure   the quality of binary classifiers and is an overall measure of accuracy at the detection of signal and non-signal classes. It is generally regarded as a balanced measure, which
can be used even if   classes have very different sizes  \cite{chicco2020,fan2009network}. Its values range from -1 to 1, and a large value indicates  good performance. 

In most cases of network construction, some links are unclassified; hence, the success
rates for the detection of existing links (SREL) and non-existing links (SRNL) are defined to study the performance of network reconstruction \cite{wang2011network,han2015robust,shi2021}, where 
\begin{equation}\footnotesize\label{eqa 26}
SREL=\frac{TP}{TP+FN+UCP},   SRNL=\frac{TN}{TN+FP+UCN},
\end{equation}
which considers the effects of non-classifiability in Table \ref{tab: illus} and is more reasonable for measuring the reconstruction accuracy of a network structure. 

To address the effect of non-classifiability on accuracy measures of reconstruction,  we define an  adjusted MCC, MCCa, by replacing FN with FN+UCP (i.e., the number of signals that are not correctly predicted) and FP with FP+UCN (the number of incorrectly predicted non-signals) in MCC. It is clear when an unclassified set disappears, MCCa reduces to MCC. It is easy to see that MCCa plays a similar role to MCC when non-classifiability occurs. We find that MCCa performs well in network reconstruction to measure the accuracy of the method from either simulations or real  examples.

\section{Numerical Studies}

We conduct   simulation studies using three kinds of model: (1) standard linear regression model with different assumptions; (2)  dynamics model of an evolutionary game   \cite{shi2021},  but with the PDG game; and (3) the Kuromato model in synchronization dynamic, which is a special case of Eq. (\ref{eq1}). Models (2) and (3) use    three   network topology structures: Erd\"os-R\'enyi (ER) random networks,  Barab\'asi-Albert (BA) scale-free networks, and  small-world Watts-Strogatz (WS) networks. In each case, we   evaluate   performance  under situations of sparse   and dense signals. 

\subsection{Linear regression models}\label{sec:NS1}

The generation model is given by
\begin{equation}\label{eqa 30}
Y={\bf{1}}_n {\bf x}_0+\Phi_1 {\bf X}_1+\Phi_2 {\bf X}_2+\epsilon
\end{equation}
where ${\bf x}_0$ is the intercept, $\bf{1}$ is an $n\times 1$ vector with all elements equal to 1, ${\bf X}_1 \in R^{p_1}$ denotes the signal parameter with elements of 1,  ${\bf X}_2 \in R^{p_2}$ denotes the non-signal parameter with elements of 0, and $\epsilon$ is the error term. A smaller $p_1$ is called a sparse signal, and a larger $p_1$ (comparing with n and $p=p_1+p_2$) is called a dense signal.  Each design matrix comes from a standard normal score with mean 0 and variance 1, but the columns   in ${\Phi}=({\Phi}_1, {\Phi}_2)$ are correlated in such a way that the correlation coefficient between $\phi_i$ and $\phi_j$ is given by $r^{|i-j|}$ with $r=0.5$  \cite{tibshirani1996regression,fan2001variable}. The error variable $\epsilon$ is generated from a Gaussian distribution with mean zero and variance $\sigma^2$. 
We also calculate the results from  several well-known shrinkage estimates, including lasso, adaptive lasso, elastic net, SCAD, MCP, and signal lasso  \cite{hastie2015book,fan2001variable,shi2021}. The abovementioned first five methods do not  shrink the parameter $X$ in the model to 1, as they were designed to shrink  irrelevant variables to zero, using different penalty functions. Hence, we call these by  lasso-type methods, while signal lasso and adaptive signal lasso are called by signal-lasso-type method in this paper.

\begin{table*}[ht]\footnotesize
	\begin{center}
		\caption{Simulation results in linear regression model based on seven methods, lasso, adaptive lasso (A-lasso), SCAD, MCP, elastic Net, signal lasso (S-lasso), adaptive signal lasso (AS-lasso). Each of the results is averaged over 200 independent realizations, where $n$ is the sample size, $p$ is the number of explanatory variables, $p_1$ is the number of signals (number of $\beta=1$). The first panel is for case of $p<n$ and signal is sparse with $ p_1=6$ . The second panel is for $p>n$  and sparse signal. The third panel consider the dense signal case, where $p_1=20$. The noise is introduced by $\sigma = 0.4, 1$, respectively, in each panel.}
		\label{tb:1} 
\setlength{\tabcolsep}{3mm}{
	\begin{tabular}{ccccccccc}	
			\hline
Method	&  \multicolumn{2}{c}{MSE/SREL/SRNL/TPR/TNR/UCR/MCC/MCCa }  \\
\hline
$(n, p, p_1, \sigma)$ 		&$(100, 30, 6, 0.4)$ & $(100, 30, 6, 1)$   \\	
\hline
lasso &0.0014/0.920/0.987/0.999/1.000/0.027/1.000/0.919 & 0.0083/0.492/0.850/0.979/0.999/0.221/0.980/0.347  \\
A-lasso &0.0015/0.925/0.976/0.999/1.000/0.343/1.000/0.900 & 0.0095/0.465/0.843/0.980/1.000/0.232/0.985/0.316 \\
SCAD & 0.0008/0.937/0.987/0.999/1.000/0.023/1.000/0.937 &  0.0043/0.522/0.985/0.999/1.000/0.108/1.000/0.632\\
MCP & 0.0008/0.935/0.992/1.000/1.000/0.019/1.000/0.943 &  0.0046/0.525/0.980/1.000/1.000/0.111/1.000/0.633 \\
ElasticNet&0.0012/0.835/0.999/1.000/1.000/0.333/1.000/0.892 & 0.0066/0.385/0.979/0.970/1.000/0.139/0.970/0.503 \\ 
S-lasso&0.0007/0.972/0.992/0.999/1.000/0.012/1.000/0.966& 0.0046/0.671/0.916/0.999/1.000/0.133/1.000/0.609 \\
AS-lasso &0.0001/1.000/0.997/0.999/1.000/0.002/1.000/0.993 & 0.0006/0.980/0.990/0.999/1.000/0.012/1.000/0.965 \\

\hline
$(n, p, p_1, \sigma)$ 	&	$(50, 150, 6, 0.4)$ & $(50, 150, 6, 1)$ \\
\hline
lasso &0.0016/0.640/0.969/0.989/1.000/0.043/0.990/0.529 & 0.0085/0.333/0.859/0.969/1.000/0.162/0.970/0.107 \\
A-lasso &0.0026/0.585/0.937/0.979/1.000/0.076/0.980/0.375 & 0.0130/0.288/0.830/0.859/1.000/0.191/0.860/0.061 \\
SCAD & 0.0002/0.822/1.000/0.999/1.000/0.007/1.000/0.896 & 0.0111/0.271/0.987/0.567/1.000/0.037/0.609/0.305 \\
MCP &0.0002/0.820/0.999/1.000/1.000/0.007/1.000/0.892 & 0.0130/0.246/0.991/0.541/1.000/0.032/0.588/0.309 \\
ElasticNet& 0.0007/0.556/0.998/1.000/1.000/0.019/1.000/0.712 & 0.0037/0.263/0.989/0.850/1.000/0.039/0.850/0.359 \\
S-lasso &  0.0004/0.806/0.994/0.999/1.000/0.013/1.000/0.839 & 0.0028/0.570/0.964/0.999/1.000/0.051/1.000/0.482 \\
AS-lasso &0.0009/0.973/0.999/0.985/0.999/0.001/0.989/0.976 & 0.0213/0.996/0.974/0.998/0.980/0.005/0.821/0.781 \\
\hline

$(n, p, p_1, \sigma)$ 	&	$(50, 150, 20, 0.4)$ & $(50, 150, 20, 1)$ \\
\hline
lasso &0.0279/0.198/0.885/0.860/0.999/0.202/0.905/0.091 & 0.0431/0.162/0.826/0.755/1.000/0.255/0.847/-0.01 \\
A-lasso &0.0104/0.429/0.925/0.964/0.999/0.140/0.975/0.372 & 0.0305/0.220/0.842/0.882/0.999/0.237/0.918/0.057 \\
SCAD & 0.1556/0.019/0.969/0.037/0.999/0.085/0.095/-0.01 &  0.1565/0.029/0.971/0.050/0.999/0.082/0.124/0.057 \\
MCP &0.1709/0.017/0.973/0.026/0.998/0.072/0.078/-0.01  & 0.1771/0.022/0.974/0.032/0.999/0.069/0.091/-0.00 \\
ElasticNet& 0.0142/0.294/0.968/0.934/1.000/0.119/0.961/0.362 & 0.0276/0.183/0.942/0.812/1.000/0.153/0.883/0.172 \\
S-lasso &  0.0015/0.850/0.972/0.999/1.000/0.044/1.000/0.815 &0.0092/0.680/0.912/0.996/0.999/0.118/0.997/0.549 \\
AS-lasso &0.0368/0.806/0.983/0.827/0.998/0.005/0.840/0.819 & 0.0400/0.794/0.982/0.812/0.998/0.004/0.827/0.809 \\
\hline

		\end{tabular}
		}
	\end{center}
\end{table*}

The results are listed in Table \ref{tb:1}, where the first panel shows the case of $p<n$ and a sparse signal, with $n=100, p=30, p_1=6$, and $\sigma = \{0.4, 1\}$. In the  second panel, $p>n$ with $n=50, p=150$, and other parameters the same as  in the first panel. The third panel considers a dense signal (or non-sparse signal), where $p_1=20$, and other parameters are the same as   in the second panel. 
The case of $p>n$ corresponds to high-dimensional variable selection and   often occurs in network reconstruction with a small number of observations. Noise is considered by adding Gaussian error with variance $0.4$ and $1$ to consider the robustness of the method.  We list all measures discussed in Section \ref{sec:3d}, but only need to check MCC, MCCa, MSE, and UCR in the table, since MCC is a synthesis  of TPR and TNR, and MCCa of SREL and SRNL. 

In the first panel, we see that all measures of accuracy from adaptive signal lasso are overwhelmingly superior to those of the other   methods, where MSE and UCR have the smallest values, and MCC and MCCa  the largest  among the seven methods. The second-best is signal lasso, which outperforms the first five  lasso-type methods in Table \ref{tb:1}  on all measures. It is seen that MCCa  of adaptive signal lasso  for high noise $(\sigma=1)$   remains  high, which indicates   robustness.  In the second panel, with $p>>n$, adaptive signal lasso   gives the largest values of MCCa and  smallest values of UCR for all cases. For small noise   $(\sigma=0.4)$, MCC of adaptive signal lasso is slightly lower,  and MSE is slight higher than for other methods, but the difference is negligible. For large noise with $\sigma=1$,  MCC of   signal lasso   is the highest, but MCCa of adaptive signal lasso is much higher than for other methods. MSE is small for all methods, but now is not smallest for AS-lasso, which  shrinks the parameter almost completely to 0 or 1,   causing a larger deviation from the true value once   falsely classified.  The last panel is for  dense signals with $p<n$, $p_1=20$, and $\sigma=0.4, 1$. As in the other panels, AS-lasso obviously outperforms the other methods  in terms of MCCa and URC. Signal lasso in this case has the largest MCC values and the smallest MSE. It is clear that the lasso-type methods (the first five) perform very poorly, especially in the case of high noise. 

In summary, we find that adaptive signal lasso outperforms the other methods and is followed by signal lasso. The first five methods (non-signal lasso) cannot  efficiently identify the correct signals, especially in cases of large noise and dense signals. Signal lasso is competitive with adaptive signal lasso only in the case of a dense signal and small noise. The most important advantage of adaptive signal lasso is that it can  classify almost all parameters to 0 or 1 (with UCR close to zero), as its theory indicates. Adaptive signal lasso is robust against noise, as  it still performs well for larger $\sigma$ in all cases.

\begin{figure}
\centering{\includegraphics[width=0.45\textwidth]{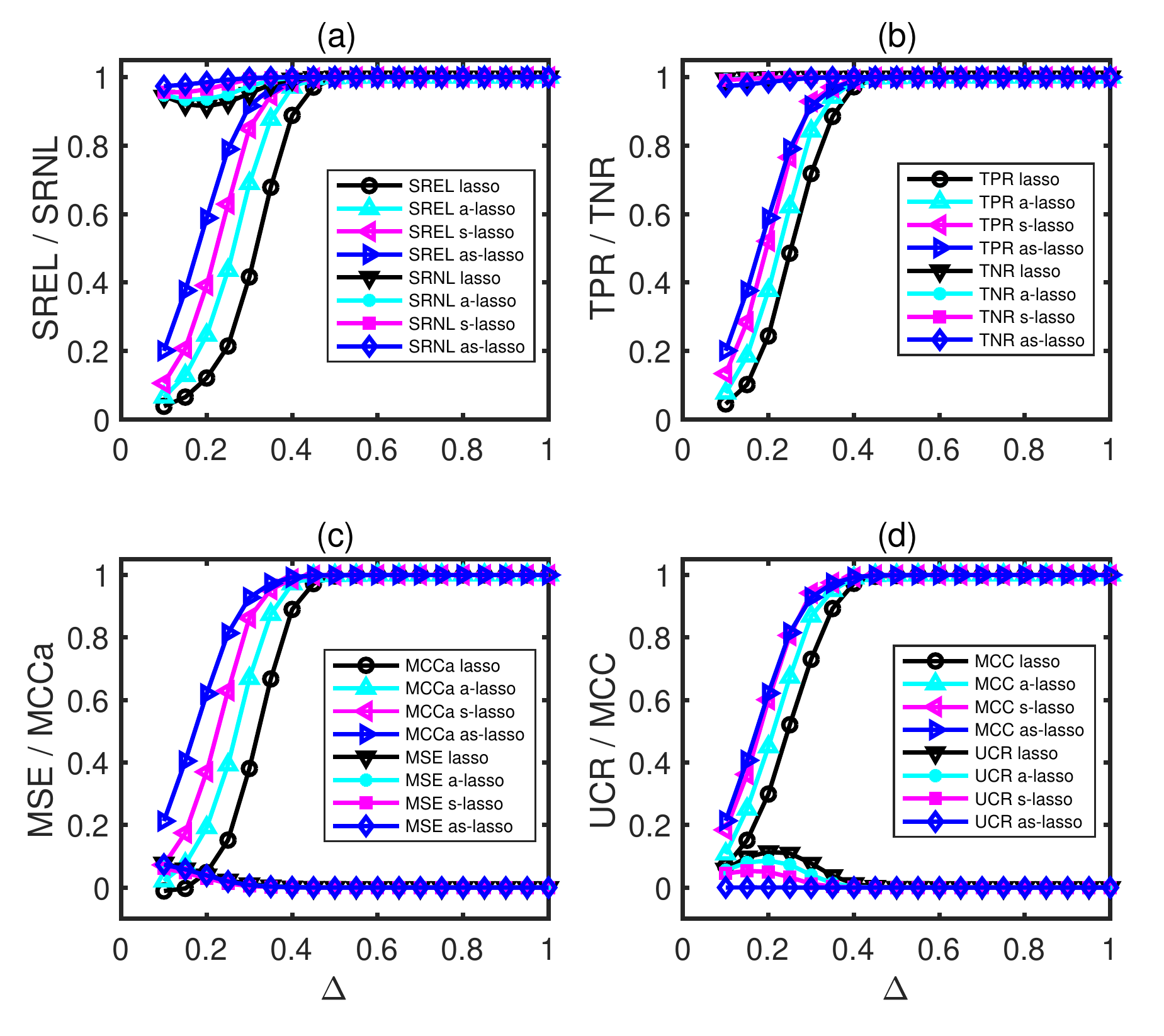}}
\caption{Accuracy in the reconstruction vs. $\Delta=L/N$, for PDG game with small world (WS) network attained by  method of lasso, adaptive lasso (a-lasso), signal lasso (s-lasso) and adaptive signal lasso (as-lasso).  (a) refer to measures using SREL and SRNL. (b) refer to measures of TPR and TNR. (c) refer to the measures of MCCa and MSE. (d)   refer to the measures of MCC and UCR. The network size $N=100$ with average degree 6. Each point is averaged over 10 simulations. }
\label{fig:3}    
\end{figure} 

\subsection{Evolutionary-game-based dynamical model}

We illustrate our method by iterative game dynamics (Eq. (\ref{eq1})) through Monte Carlo simulation.  We take a simple structure with $R=1,T=b=1.15$, and $P=S=0$ in our simulation. The total payoff is $F_i=\sum_{j} a_{ij} P_{ij}$ (eq. (\ref{eq3})), where $a_{ij}=1$ if the $i$th player and $j$th player interact, and 0 otherwise. In each round of the game, each player   calculates its total payoff and then imitates with a certain probability the two strategies (p, q) of a randomly selected player in its direct neighborhood. That is, player $x$ adopts the strategy of player $y$ with probability $W=1/\{1+\exp{[(F_x-F_y)/K]}\}$ (Szabo, {\it et al.}, 2007), where $K$ is the uncertainty in   strategy transition,   with a value of 0.1 in this paper. 
The game  iterates forward in a Monte Carlo manner and player $i$ (the focal player) acquires its fitness (total payoff) $F_i$ by playing the game with all its direct neighbors, i.e., $F_i=\sum_{j=1}^N a_{ij} P_{ij}$. The focal player then randomly picks a neighbor $j$, which similarly acquires its fitness. Following the definition of fitness, a strategy update then occurs between its direct neighbors in a given network. Player $i$ tries to imitate the strategy of player $j$ with  Fermi updating probability $W=1/(1+\exp [(F_i-F_j)/K])$, where $K=0.1$  \cite{szabo1998evolutionary,szabo2002evolutionary}. To make the model more realistic, we   account for  mutation at very small rates.  

Now $F_i=\sum_{j=1, j \neq i}^N a_{ij} P_{ij}$ can be written as a linear regression model 
\begin{equation}\label{eqa29}
Y_i=\Phi_i \tilde X_i +e_i,
\end{equation}
where $Y_i=(F_i(t_1), F_i(t_2), \cdots, F_i(t_L))' $, $\tilde X_i=(a_{i1}, \cdots, a_{iN})'$, and
$\Phi_i$  has the form of 
\begin{displaymath}
\left(\begin{array}{cccccc}
P_{i1}(t_1) & \cdots & P_{i,i-1}(t_1) & P_{i,i+1}(t_1) & \cdots & P_{i_N}(t_1)\\
P_{i1}(t_2) & \cdots & P_{i,i-1}(t_2) & P_{i,i+1}(t_2) & \cdots & P_{i_N}(t_2)\\
\vdots & \vdots & \vdots & \vdots      \\
P_{i1}(t_L) & \cdots & P_{i,i-1}(t_L) & P_{i,i+1}(t_L) & \cdots & P_{i_N}(t_L)
\end{array}\right), 
\end{displaymath}
Let $Y=(Y'_1, \cdots, Y'_N)'$, $X=(\tilde X'_1, \cdots, \tilde X'_N)'$, $\Phi=diag(\Phi_1, \Phi_2, \cdots, \Phi_N)$, then  Eq. (\ref{eqa29}) can be converted into the general form of Eq. (\ref{eqa01}). 

\begin{figure*} 
\centering{\includegraphics[width=0.7\textwidth]{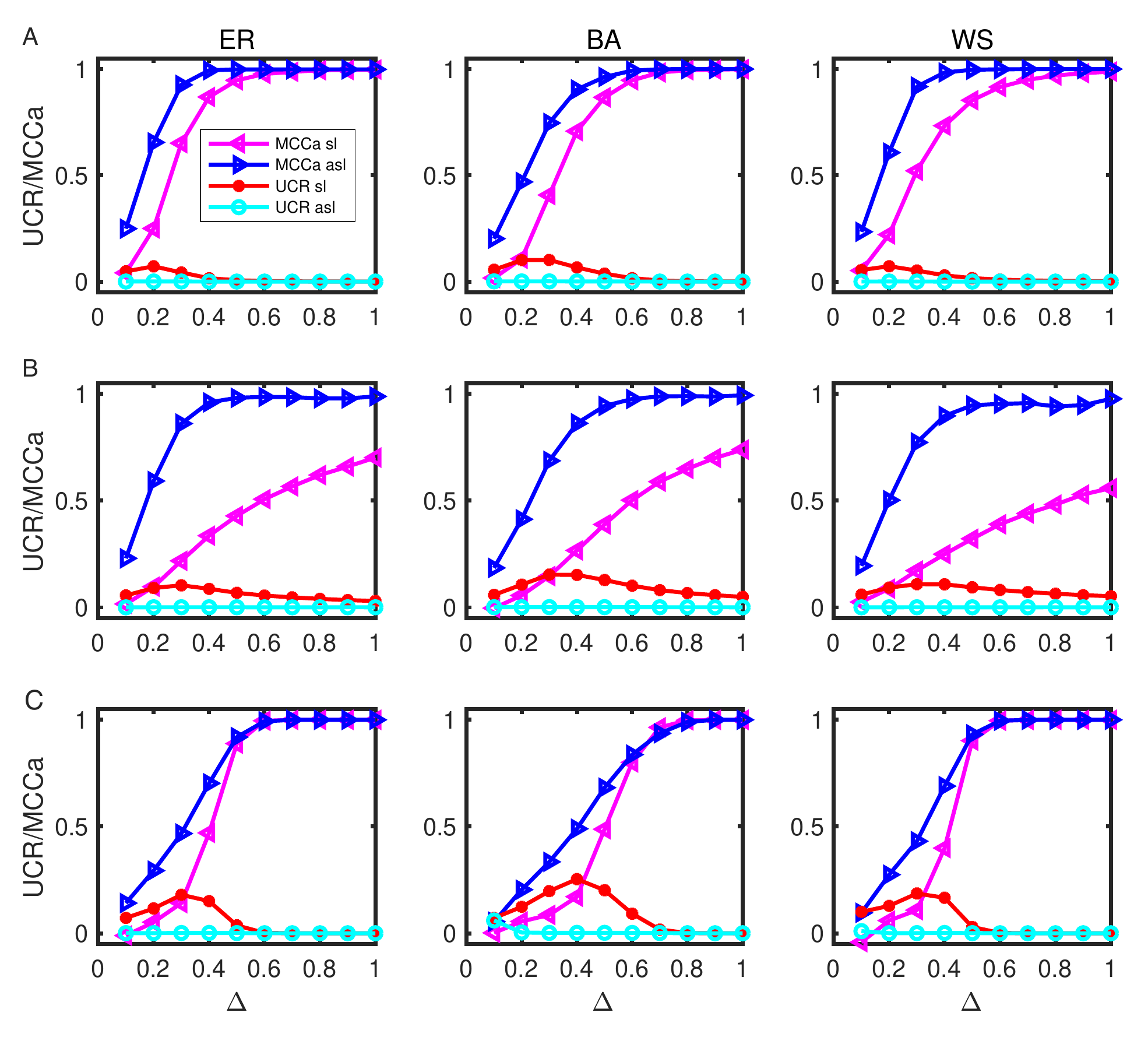}}
\caption{Accuracy measures MCCa and UCR in the reconstruction vs. $\Delta=L/N$, for PDG model attained by  method of signal lasso (sl) and adaptive signal lasso (asl) in three kinds of network.  The panel A refers to results for PDG game with noise $\sigma^2=0.1$ and average degree 6. The panel B refers to the results for PDG game with noise $\sigma^2=0.3$ and average degree 6. The panel C refers to the results for PDG game with average degree 20 without noise. Three columns correspond to the results based on Erd\"os-R\'enyi (ER) random network,  Barab\'asi-Albert (BA) scale-free network and  small world (WS) network, respectively. The network size $N=100$ and each point is averaged over 10 simulations.  }
\label{fig:4}    
\end{figure*}

Fig.  \ref{fig:3}  plots the measures   in Section
\ref{sec:3d}  against the data length  $\Delta=L/N$ in the PDG
model with a small world (WS) network, using the methods
of lasso, adaptive lasso, signal lasso,
and adaptive signal lasso, where adaptive lasso
  represents one that performs best among the lasso-type methods
and lasso is also included as a commonly used
method (results for all methods are given in Supplemental Material and show that
adaptive signal lasso performs best). It is clear that adaptive signal lasso performed
best based on most   measures, followed by signal
lasso, adaptive lasso, and lasso. For measures TNR and
MSE, adaptive signal lasso is not absolutely superior, but the differences of four methods are very small.
It is of interest to see that UCR of adaptive signal lasso is close to zero even in the case of small  $\Delta=L/N$,
which is an appealing property. 

Fig.  \ref{fig:4}  A and B show the reconstruction accuracy measures
of   MCCa and UCR in the ER, WS, and BA networks
respectively for $N = 100$ in the PDG model when a noise is added
to the model. We only list the results for signal lasso and
adaptive signal lasso, since other methods performed more poorly
than these two methods. Adaptive signal
lasso is obviously superior to signal lasso   when there is
noise in the data, and UCR   remains best (close to
zero). For the large variance $\sigma^2 = 0.3$, adaptive signal lasso
can still resist effect of noise according to Fig.  \ref{fig:3}, which indicates the robustness of adaptive signal lasso.
 Fig.  \ref{fig:4}C shows the results of a dense
network with average degree equal to 20 and shows that
adaptive signal lasso has higher values of   MCCa than signal lasso at a small $\Delta=L/N$ 
and approach being equal at a large $\Delta=L/N$ ,
however, the UCR of adaptive signal lasso is always better than that of signal lasso.

\subsection{\label{sec4} Kuramoto model in synchronization problem}

For problems introduced in Eq. (\ref{eq1}), we use the Kuramoto model  \cite{boccaletti2014structure,timme2007revealing,wu2012inferring} to illustrate the reconstruction of the network in a complex system. This model has the following governing equation:
\begin{equation}\label{eq10}
\frac{d\theta_i}{dt}=\omega_i +c\sum_{j=1}^N a_{ij} sin(\theta_j -\theta_i),
\end{equation}
$i=1,\cdots, N$, where the system is composed of $N$ oscillators with phase $\theta_i$ and coupling strength $c$, each of the oscillators has its own intrinsic natural frequency $\omega_i$, $a_{ij}$ is the adjacency matrix of a give network and is need to be estimated in network reconstruction. Using the same framework of reference\cite{shi2021}, the Euler method can be employed to generate time series with an equal time step $h$.  Let $Y_i=(y_{i1}, \cdots, y_{iL})'$, $y_{it}=[\theta_i(t+h)-\theta_i(t)]/h$, $\Phi_i=(\phi_{ij}(t))$ is a $L\times N$ matrix\cite{shi2021}, with elements

$$\phi_{ij}(t)=c \times sin(\theta_j (t)-\theta_i(t))$$
for $t=1,\cdots,L$ and $j=1,\cdots,N$, $\tilde X_i=(a_{i1}, \cdots, a_{iN})'$, then reconstruction model can be written as
\begin{equation}\label{eq13}
Y_i=\omega_i {\bf 1}_L+\Phi_i \tilde X_i, 
\end{equation}
where ${\bf 1}_L$ denote a $L\times 1$ vector with all element 1.

\begin{figure*} 
\centering{\includegraphics[width=0.7\textwidth]{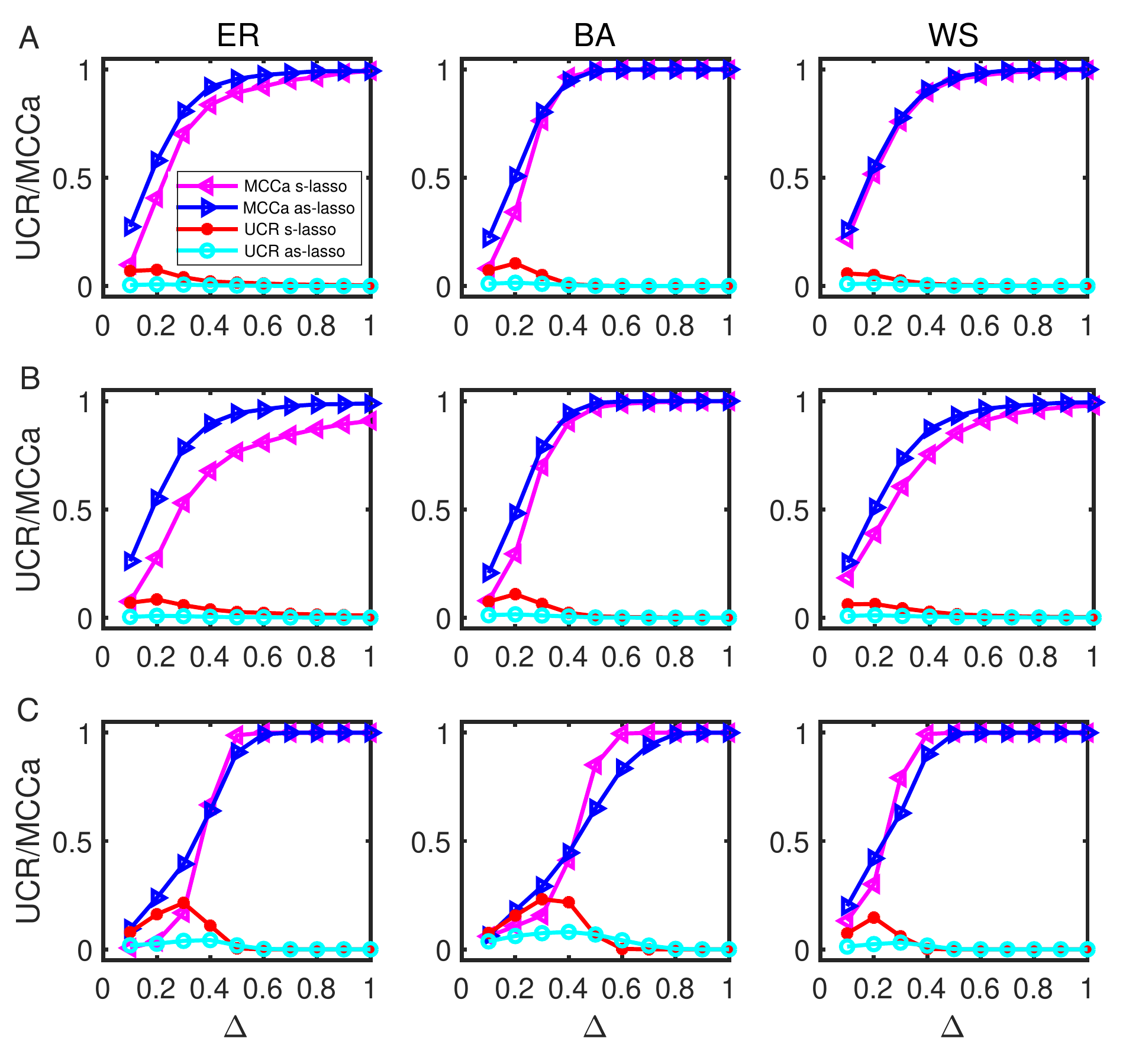}}
\caption{Accuracy measures MCCa and UCR in the reconstruction vs. $\Delta=L/N$, for Kuramoto model attained by  method of signal lasso (sl) and adaptive signal lasso (asl) in three kinds of network.  The top panel contains the measures from the Kuramoto model in the network with average
6, while middle panel contains the results for the case with
noise of $\sigma^2 = 0.3$, and the network with average 6. The bottom
panel gives the results for dense network with average degree
20 and without noise. Three columns correspond to the results
based on Erd\"os-R\'enyi (ER) random networks,  Barab\'asi-Albert (BA) scale-free networks and  small world (WS) network,
respectively. The network size $N = 100$ and coupling
c=10 for all cases. Each point is averaged over 10 simulations. }
\label{fig:5}    
\end{figure*}

Fig. \ref{fig:5}  consider the Kuramoto model in ER, WS and
BA networks respectively for $N = 100$ and coupling
strength $c=10$. Panel A is the results without the noise,
where we find the similar results as before analysis. Panel
B is the result with variance of noise equal to 0.3, and it
also show adaptive signal lasso method is robust against the noise. Panel
C give the result of dense network with average degree 20,
it is interest to observe that adaptive signal lasso performed better than
signal lasso for small $\Delta=L/N$, while signal lasso become better than
adaptive signal lasso for large $\Delta=L/N$, where UCR always performed
well. This results implies that signal lasso is also useful
in dense network, however adaptive signal lasso is more computationally
convenient since it only need tuning one parameter.

\section{Examples}

\subsection{Human behavioral data}

\begin{figure} 
\centering{\includegraphics[width=0.47\textwidth]{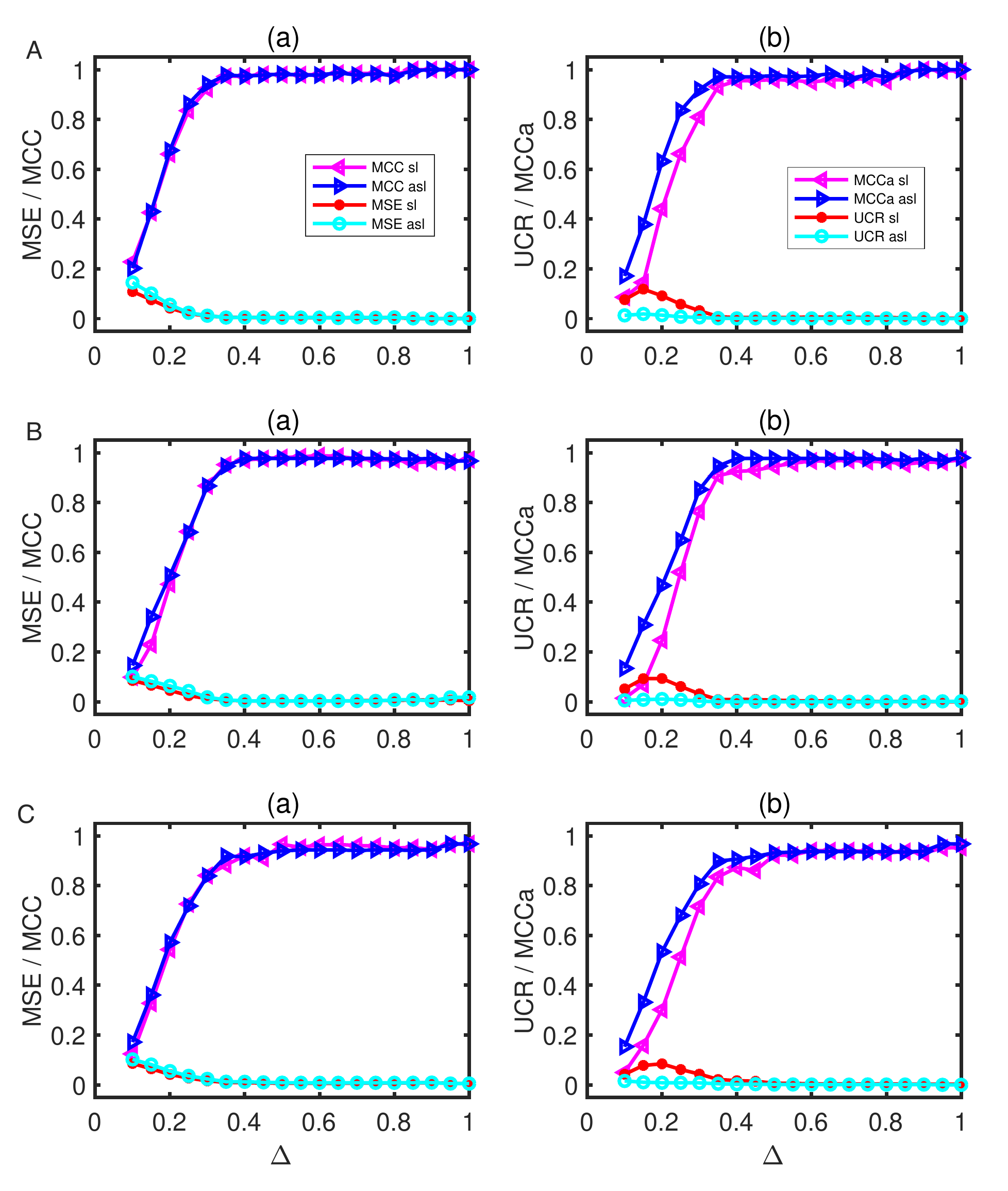}}
\caption{Accuracy in the reconstruction vs. $\Delta=L/N$, for three real trials using  method of signal lasso (sl) and adaptive signal lasso (asl) .  Panel A refers to the results of the experimental ring network, where Panels A(a) is plot of MCC and MSE vs. $\Delta$, A(b) refer to the MCCa and UCR criterion, for signal lasso and adaptive signal lasso, respectively. There are 35 nodes and 140 links and the degree of each node is 4. Panel B refers to the results of the experimental homogeneous random network. There are 50 nodes and 200 links and the degree of each node is 4. Panel C refers to the results of the experimental heterogeneous random network. There are 50 nodes and 200 links and the average degree of each node is 4.  }
\label{fig:6}    
\end{figure}

We  use  real data from a human behavior experiment, whose   purpose   is to study the impact of   punishment on network reciprocity, to illustrate   social network reconstruction \cite{li2018punishment}. A total of 135 volunteers from Yunnan University of Finance and Economics and Tianjin University of Finance Economics participated in experiments with three trials. In treatment I, 35 participants played an iterative prisoner's dilemma game with punishment on the static ring network with four neighbors. In treatments II and III,   50 participants  participated in the same experiment, where treatment II was implemented on a homogeneous random network with degree   4, and in treatment III, each player was placed on a heterogeneous random network, where half the nodes had degree 3, and  half had degree 5. The payoff matrices are presented in Table \ref{t3}.  Each player played with   direct neighbors in each round to gain their payoff and updated the strategy to optimize their future payoff. The number of interactions in each session was 50, which was unknown to players. 
This dataset was analyzed to show that the signal lasso method outperformed the lasso and CS methods \cite{shi2021}; here we use this example again to study the performance of proposed method.  The results are summarized in   Fig. \ref{fig:6}, where panels A, B, and C refer to the results of the experimental ring,  homogeneous random, and heterogeneous random network, respectively. The figures in column (a) list the measures of MCC and MSE, and column (b) lists the measures of MCCa and UCR. It is clear that both for MCC and MCCa, adaptive signal lasso is superior to signal lasso, especially for MCCa. MSE of adaptive signal lasso is slightly larger than that of signal lasso for smaller values of $\Delta$. The UCR of adaptive signal lasso is close to zero for all values   of $\Delta$, as   expected, but UCR of signal lasso shows that it still contains some unclassified parts for smaller $\Delta$, which decreases reconstruction accuracy.

\begin{table}[!htb]
	\caption{The payoff matrix of PDG with punishment option (treatment I) and the standard PDG (treatment II and treatment III) are shown in left panel and right panel, respectively.}
    \label{t3}
	\begin{minipage}{.5\linewidth}
		\centering
		\setlength{\tabcolsep}{4mm}{
			\begin{tabular}{cccc}
				& C & D & P \\
				\hline
				C & 2 & -2 & -5 \\
				D & 4 & 0 & -3  \\
				P & 2 & -2 & -5  \\
				\hline
		\end{tabular}}
	\end{minipage}%
	\begin{minipage}{.5\linewidth}
		\centering
		\setlength{\tabcolsep}{4mm}{
			\begin{tabular}{ccc}
				~ & C & D  \\
				\hline
				& & \\
				C & 4 & -2  \\
				D & 6 & 0   \\
				\hline
		\end{tabular}}
	\end{minipage}
\end{table}

\subsection{World Trade Web (WTW)}

A network formed by import/export relationships between
  countries, the World Trade Web
(WTW) has been extensively studied. Some empirical
studies focused on the WTW as a complex network and
investigated its architecture \cite{squartini2011}. When the available information on the system is incomplete or partial,   reconstruction methods of the
whole network have been proposed, such as maximum entropy
  \cite{Mistrulli2011,squartini2018}
and the configuration model (CM)  \cite{Garlaschelli2004,Garlaschelli2009,Garlaschelli2013}. Much of the network
reconstruction research in WTW focuses   on  
ensemble models, which means that a model is defined
to be not a single network, but a probability distribution
over many possible networks \cite{park2004}. We use a new perspective to illustrate the construction of a trade network using our proposed signal
lasso method, which gives a model-based estimation
of adjacency matrix A.

\begin{table} 
\caption{Estimation accuracy of reconstruction  of  trade network }
\label{tab:real1}
\begin{center}
\setlength{\tabcolsep}{8pt}
\begin{tabular}{lcccc} 
\toprule 
Measures&lasso&A-lasso&S-lasso& AS-lasso  \\
 \hline
 SREL & 0.0055& 0.0202 &0.9632& 0.9710 \\
SRNL  &1 &1 &1 &1 \\
TPR    &0.0075 &0.0263 &0.9652 &0.9722 \\
TNR    &1 &1 &1&1 \\
MCC   &0.0246 &0.0563 &0.9659 &0.9722 \\
MCCa &0.0185 &0.0446&0.9636 &0.9282 \\
UCR   &0.2296 &0.2038 &0.0103 &0.0022 \\
MSE   &1.3e+08 &1.9e+03 &0.266  &0.1358 \\
\hline
\end{tabular}
\end{center}
\end{table}

The database of the trade network is built from the data
directly reported by each country to the United Nations
Statistical Division, which is freely available  from
their website, which provides disaggregated data
on bilateral trade 
flows for more than 5000 products
and 200 countries from 1995 to 2018. Each trade  flow is characterized by a combination of exporter-importer-products-
year and provides the value and   quantity
of the flow. Because some countries do not report the values of their exports or imports, or because a large amount of data are missing, we  only
used data from 215 countries in the years   1995--2018, realizing a network of all   trade products with 215
nodes and 36296 links, which is   extremely dense   \cite{squartini2011}.

We only considered an un-directional binary network, defining two countries  as connected if one 
has output to another. Let $Y_{it}$ denote total foreign trade (in U.S. dollars) of the $i$th country,
  including total imports and exports with other countries.
Let $w_{ij}(t)(= w_{ji}(t))$ denote the import and export
values of the $i$th country with the $j$th country at time
$t (t=1,\cdots, T, T=24)$. Then, it is obvious that
\begin{equation}\label{eq15}
Y_{it}=\sum_{j=1}^N a_{ij}w_{ij}(t)+\epsilon_{it}
\end{equation}
where $a_{ij}$ is an element of a connectivity
matrix with $a_{ii}=0, a_{ij}=a_{ji}$ for $i \ne j$, and $N=215$ is the number of   countries. It is noted that for some country pairs, the export (or import) 
bilaterally or unilaterally at some time periods are missing,
that means in some years during the observations there
are no trade between two countries or missing because
the trading quantity is small and can be ignored. Hence,
the   edges during 24 years might be incompletely  linked, which is  unbalanced.

In many economic network analyses and spatial panel
data models, the adjacency matrix must be predetermined. A simple method to obtain an adjacency
matrix is to define $a_{ij}=1$ if there is a connection
at least one time between two nodes, and $a_{ij}=0$ otherwise
 \cite{anslin2013,zhu_xn2019}. However, this may be unreasonable
if two countries only have one year of very small or negligible trade. An alternative method is
to use the percentage of the trade between two
countries of the total trade amount to measure their
connectivity, but they need a cutoff value, which might
be subjective. 
An interesting problem is determining how to use the structural relationship, such as Eq. (\ref{eq15}), to estimate the network during the observed
periods; this problem  can be dealt
with using the shrinkage theory proposed in this paper
since in this case  $a_{ij}$ is either 0 or 1.
It is noted that the model in Eq. (\ref{eq15}) relates to the reconstruction of binary topology, and we assume this   is fixed during the observed time period, which is usually favorable in statistical network models \cite{anslin2013,zhu_xn2019}.

For comparison, we define the reference connectivity
as $\tilde a_{ij}$ =1 if at least one   connection occurs during
the observed years, and zero otherwise. This   is not assumed to be a correct adjacency
matrix, but it can be used for comparison and  
analysis. The results are listed in Table \ref{tab:real1}, using
$\tilde a_{ij}$ as a baseline for comparison. It is surprising that lasso
and adaptive lasso   perform   poorly in terms
of SREL (TPR), MCC, MCCa, and MSE,
even with high values of SRNL (TNR), perhaps because the trade network is  highly dense, which  was verified in our simulation
of Section \ref{sec:NS1}. The performance of signal lasso
and adaptive signal lasso is excellent, and they
give   similar results. All methods can correctly identify non-existent links (SRNL and TPR have values of 1) because there are just a few zero edges  (about 12\%) in WTW. 
The values of SREL (TNR) of adaptive signal
lasso can exceed 96\%, even   with $\Delta= 24/215 = 0.11$.

\begin{figure}
\centering{\includegraphics[width=0.5\textwidth]{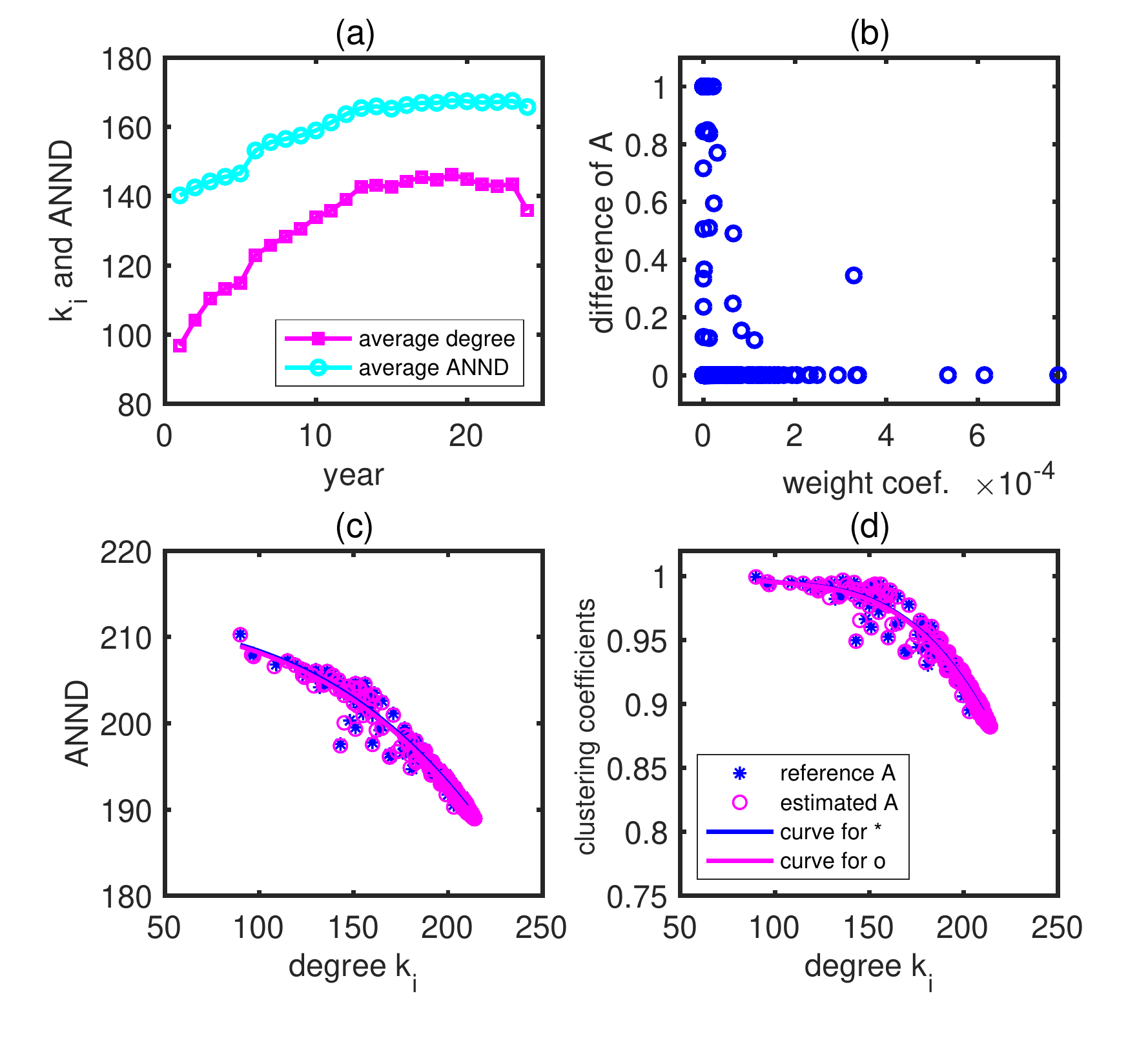}}
\caption{The reconstruction results for undirected WTW using
method of adaptive signal lasso method. (a) plots of average
degree and average nearest neighbor degree (ANND) versus
year, where 1 represent 1995 and 24 for 2018. (b) plot of diference
of reference and estimated adjacency matrix of network.
(c) plot of average nearest neighbor degree (ANND) versus
average degree. (d) plot of clustering coefficients of network
versus average degree. In (c) and (d) curves are fitted based
on polynomial of order 3.}
\label{fig:7}    
\end{figure}

Fig.  \ref{fig:7} shows some basic statistics calculated from the reference
and estimated adjacency matrix. Fig.   \ref{fig:7}(a) shows the
evolution of the average degree and average nearest neighbor
degree (ANND) over the time calculated from $\tilde a_{ij}$, which
shows that the trade network is time-dependent and has increasing
degrees of nodes, but there is a downward trend
in the last few years. Fig.  \ref{fig:7}(b) shows the absolute difference
of reference adjacency matrix $\tilde A$ and estimated adjacency
matrix $\hat A$ (denoted by $D(\tilde A;\hat A) =
\vert \tilde a_{ij}- \hat a_{ij}\vert$) against
 their weight coefficients
$$
\tilde{w}_{ij}=\frac{\sum_{t}w_{ij}(t)}{\sum_t \sum_{i<j}w_{ij}(t)}.
$$
The small value of $\tilde w_{ij}$ means small imports/exports between two countries
or few connections during 24 years. $D(\tilde A, \hat A) =1 (0)$
means inconsistent (consistent) results between $\tilde A$ and $\hat A$, and
$0< D(\tilde A;\hat A) <1$ indicates the unclassified cases. It is
clear that inconsistency only occurs at very small weight
coefficients, and unclassified cases occur mostly at
smaller weight coefficients. Further investigation finds that
these two cases occur at $J = 1, 2, 3$, where $J$ is
the number of years from 1995 to 2018 for which import/export statistics are available (namely nonzero). These results
show that shrinkage estimation can eliminate  
unimportant linkages between nodes. 

Fig.  \ref{fig:7}(c) and (d) show plots of average nearest neighbor degree
(ANND) and clustering coefficients,
respectively, versus average degree. We find the results from $\tilde A$ and $\hat A$ are
highly coincident, where decreasing trends have been
found in previous studies employing different datasets in
WTW  \cite{squartini2011}. The decreasing
trend in Fig.  \ref{fig:7}(c) is known as disassortativity (i.e., countries trading with highly connected countries
have  few trade partners, and those trading
with poorly connected countries have many) in WTW \cite{squartini2011}.

\section{Conclusions}

We proposed   adaptive
signal lasso, based on the signal lasso method, to estimate
  signal parameters and uncover the   topology
in complex networks, adding a weight to penalty terms of signal lasso. The theoretical
properties of this method were studied, and  
simulations were conducted, which included linear
regression models, an evolutionary game dynamic model,
and a simple synchronization model. Two real examples
were analyzed. The results showed that our
  method can effectively uncover   signals in
network reconstruction in complex systems. Compared
with signal lasso and other   lasso-type methods,
adaptive signal lasso has three advantages: (1) It only needs the tuning of one parameter in a
small range in (0,1), which greatly reduces the computational
cost; (2) Adaptive
signal lasso is more robust to noise and contamination,
as   shown in the simulation of the PDG evolutionary
model and Karumoto model, where adaptive signal
lasso   maintains good performance if the variance
is not too large, while  other methods fail; (3) Signal-type lasso, including signal lasso and adaptive
signal lasso, outperforms  other lasso-type methods, which is useful for the   reconstruction of 
some non-sparse networks.

In the reconstruction of a WTW network, we   used a
simple example to illustrate our method, since a reference
network is usually used in practice. We found our proposed
method to be robust and efficient for both sparse and dense
networks. When a directed network was considered, a similar
analysis was conducted, and we also found that the signal-lasso-type method   performed  very well (see Supplemental Material). In practical
situations, only partial information is available; e.g., we might
only know  in-degree or out-degree values in a binary network, in which case the problem can be solved by replacing $P_{ij}$ in eq. (\ref{eq15}) using an existing estimation method \cite{squartini2018}. 

We only considered the detection of signal parameters (0 or 1) in binary networks. In weighted networks, the weight coefficient belongs to the interval [0,1] \cite{squartini2018,wang2011network,Garlaschelli2013}. In such a circumstance, the proposed adaptive signal lasso  can be applied by tuning the parameter $\lambda$ in Eq. (\ref{eqa16}) to an appropriate value instead of letting it go to infinity (taking a large enough value as in our problem here). This topic will be studied in our future research.

\section*{Acknowledgement}
This work was supported by NSFC under Grant No. 11931015, 12271471 to L. S. and JSPS Postdoctoral Fellowship Program for Foreign Researchers (No. P21374) and an accompanying Grant-in-Aid for Scientific Research to C.\,S.

\vspace{5mm}

\section*{Appendix}

\renewcommand{\theequation}{A.\arabic{equation}}

\subsection{Comparison of weight  choice}
\begin{figure}
\centering{\includegraphics[width=0.48\textwidth]{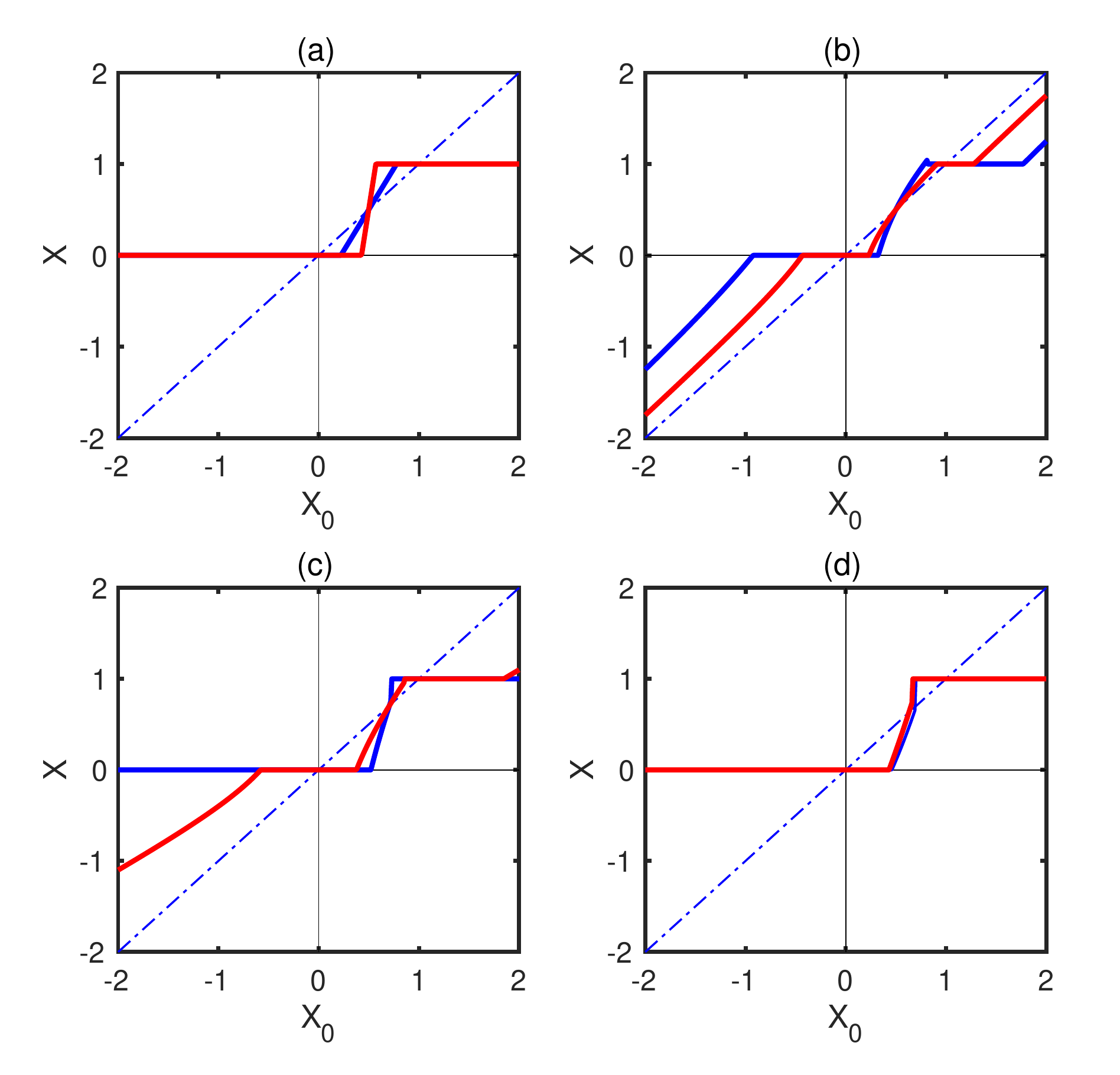}}

\caption{Solution of $X$ under   orthogonal design in adaptive signal lasso with different weights. (a) solution of adaptive signal lasso as function OLS $\hat X_0$ with $\omega_{1j}=1$ and  $\omega_{2j}=|\hat X_{j0}|$, where red line is for $\lambda_1=2$ and $\lambda_2=4$, blue line is for  $\lambda_1=0.4$ and $\lambda_2=0.8$; (b) solution of adaptive signal lasso as function OLS $\hat X_0$ with $\omega_{1j}=|\hat X_{j0}|^{-1}$ and  $\omega_{2j}=|\hat X_{j0}|$, where red line is for $\lambda_1=0.2$ and $\lambda_2=0.8$, blue line is for  $\lambda_1=0.6$ and $\lambda_2=1.2$; (c) solution of adaptive signal lasso as function OLS $\hat X_0$ with $\omega_{1j}=|\hat X_{j0}|^{-1}$ and  $\omega_{2j}=1$, where red line is for $\lambda_1=0.1$ and $\lambda_2=0.2$, blue line is for  $\lambda_1=0.3$ and $\lambda_2=0.6$; (d) solution of adaptive signal lasso as function OLS $\hat X_0$ with $\omega_{1j}=1$ and  $\omega_{2j}=|\hat X_{j0}|^{\gamma}$, with  $\lambda_1=2$ and $\lambda_2=4$, blue line is for  $\gamma=1.5$, and $\gamma=2$.}
\label{fig:a1}       % Give a unique label
\end{figure}

If we  assign the weight in both penalty terms, then  $\omega_{1i}=\vert \hat X_0\vert^{-\nu}$ and $\omega_{2i}=\vert \hat X_{i0}\vert^\gamma$ are two appropriate choices in adaptive signal lasso, where $\hat X_{i0} $ is the OLS estimator of $X$. Fig.  (\ref{fig:a1})(a) shows the curves of the penalty solution of   adaptive signal lasso as a function of $\hat X_0$ when $\nu=0,\gamma=1$, which is  the formula   used throughout this paper. Fig.  (\ref{fig:a1})(b)--(d) show the solution of   adaptive signal lasso as a function of $\hat X_0$ in the three cases of  $\nu=1$ and $\gamma=1$, $\nu=0$ and $\gamma=1$, and $\nu=0$ and $\gamma=1.5, 2$, respectively. Although adaptive signal lasso in (b)--(d) also has the function of shrinking the values between 0 and 1 in two directions (0 or 1), it is too tedious and complicated, and sometime is unstable. Case (a), with $\nu=0$ and $\gamma=1$ is   the simplest but it can achieve our purpose very well. In addition, case (a) can reveal an appealing property  (such as the result in Eq. \ref{eqa17})  in selecting the tuning parameters in adaptive signal lasso, as we show  in Eq. (\ref{eqa17}). 

\subsection{The proof of Eq. (\ref{eqa15}) and (\ref{eqa17})}

In order to study the geometry of signal lasso, we assume that the columns of $\Phi$ are orthogonal each other and $p < n$. The ordinary least squares estimate in this special case then has the form of $\hat X_0=\Phi' Y$. Let $\hat{Y}_0=\Phi \hat X_0$, we then have
\begin{equation}\label{ap01}
\begin{array}{rl}
L(X, \lambda_1,&\lambda_2)
=\frac{1}{2} \Vert Y-\hat Y_0\Vert^2_2+\frac{1}{2} \Vert X-\hat X_0\Vert^2_2+\\
&\lambda_1 \Vert {\omega}_1\circ X\Vert_1+\lambda_2\Vert {\omega}_2\circ (X-\mathbf{1}_p)\Vert_1,
\end{array}
\end{equation}
where ${\omega}_1$ and ${\omega}_2$ are $p\times 1$ known weight vectors
and $\circ$ denotes the Hadamard production.
Note that the first term is constant with respect to $X$ and $\Vert X (X-\hat X_0)\Vert_2^2=\Vert (X-\hat X_0)\Vert_2^2$. Using the fact that
\begin{equation}\label{ap02}
%\begin{split}
\frac{\partial \vert X_i \vert}{\partial X_i}=\left\{
\begin{array}{l}
1, \ \ \ \ \ \ \ if \ X_i>0, \\
-1, \ \ \ \ \ \ \ if \ X_i < 0, \\
\in \ [-1,1] \ \ if \ X_i=0,
\end{array}\right.
%\end{split}
\end{equation}
and differentiating $L(X, \lambda_1,\lambda_2)$ with respect to $X$
and set it to zero, after some calculations we have
\begin{equation}\label{ap03}
%\begin{split}
\hat X_j=\left\{
\begin{array}{l}
(\hat X_{j0}+\lambda_1 \omega_{1j}+\lambda_2\omega_{2j})_{-}, \ \ \ \ \ \ \ \hat X_{j0}\leq0, \\
(\hat X_{j0}-\lambda_1\omega_{1j}+\lambda_2 \omega_{2j})_{+}, \ \ \ \ \ \ \ 0<\hat X_{j0}\leq \delta, \\
\max\{1, \hat X_{j0}-\lambda_1\omega_{1j}-\lambda_2\omega_{2j} \}, \ \ \hat X_{j0}> \delta,
\end{array}
\right.
%\end{split}
\end{equation}
in which $\delta$ is the solution of function $\hat X_{j0}-\lambda_1 \omega_{1j}+\lambda_2 \omega_{2j}=1$.   From our analysis in the main content of this paper, we choose that $\omega_{1j}=|\hat X_{j0}|^{-\nu}$ as adaptive lasso used and $\omega_{2j}=|\hat X_{j0}|^{\gamma}$.  When $\nu=0$ and $ \gamma=1$, Eq. (\ref{eqa15}) can be derived easily from Eq. (\ref{ap03}) .

Now we re-parameterize $\lambda_1$ and $\lambda_2$ by $\lambda$ and $\alpha$, then we have
\begin{equation}\label{ap04}
%\begin{split}
\hat X_j=\left\{
\begin{array}{l}
\{\hat X_{j0}+\lambda (\alpha- \hat X_{j0}) \}_{-}, \ \ \ \ \ \ \ \hat X_{j0}\leq0, \\
\{\hat X_{j0}-\lambda (\alpha-\hat X_{j0})\}_{+}, \ \ \ \ \ \ \ 0<\hat X_{j0}\leq \alpha_2, \\
\max\{1, \hat X_{j0}-\lambda (\alpha+\hat X_{j0})\}, \ \ \hat X_{j0}> \alpha_2,
\end{array}
\right.
%\end{split}
\end{equation}
 which will immediately lead to the results in Eq. (\ref{eqa17}).

\subsection{Coordinate descent algorithm}

Differentiating (\ref{eqa12}) with respect to $X_k$ and equal it to zero, we have
\begin{equation}\label{ap05}
-\sum_{i=1}^n (y_i-\sum_{j=1}^p \phi_{ij}X_j)\phi_{ik}+\lambda_{1k}^* s_k^{(1)} + \lambda_{2k}^* s_k^{(2)}=0, \\
\end{equation}
where $s_k^{(1)}=\partial \vert X_k \vert/\partial X_k=sgn(X_k)$, $s_k^{(2)}=\partial \vert X_k -1\vert/\partial X_k=sgn(X_k-1)$; sgn(z) takes values of sign(z) for $z\neq 0$ and some value lying in $[-1,1]$ for $z=0$. Let $r^{(k)}=(r^{(k)}_1, \cdots, r^{(k)}_n)' $  denote the partial residual, where $r^{(k)}_i=y_i-\sum_{j\neq k} \phi_{ij}X_j$, then using formula (\ref{ap02}) and after some calculation, we have

\begin{equation}\label{ap06}
 X_k=\left\{
\begin{array}{l}
\displaystyle [z_k+\delta_{1k}^* ]_{-}, \ \ \ \ \ \ \ \ z_k\leq 0, \\
\displaystyle [z_k-\delta_{2k}^* ]_{+}, \ \ \ \ \ \ \ 0<z_k \ \leq 1+\delta_{2k}^*, \\
\displaystyle\max\left\{1, [z_k-\delta_{1k}^*] \right\}, \ \ z_k \ > 1+\delta_{2k}^*,
\end{array}\right.
\end{equation}
where $\left< z, y\right>$ denote the inner product of vectors $z$ and $y$, $\delta_{1k}^*=(\lambda_{1k}^*+\lambda_{2k}^*)/\left< \phi_k, \phi_k\right>$ and  $\delta_{2k}^*=(\lambda_{1k}^*-\lambda_{2k}^*)/\left< \phi_k, \phi_k\right>$, and  $z_k= \left< r^{(k)}, \phi_k\right>/\left< \phi_k, \phi_k\right>$. 
From the definition of $S_{\theta_1, \theta_2}(z)$, it is easy to see
\begin{equation}
X_k=S_{\delta_{1k}^*, \delta_{2k}^*}\left( \displaystyle\frac{\left< r^{(k)}, \phi_k\right>}{\left< \phi_k, \phi_k\right>}\right),
\end{equation}
Note that
$\left<r^{(k)}, \phi_k\right>=\left< r, \phi_k\right>+X_k \left< \phi_k, \phi_k\right>$,  where $r=Y-\Phi X$, we have
\begin{equation}\label{ap10}
X_k=S_{\delta_{1k}^*,\delta_{2k}^*}\left( X_k+\displaystyle\frac{\left< r, \phi_k\right>}{\left< \phi_k, \phi_k\right>}\right)
\end{equation}
Therefore the update can be written as
\begin{equation}\label{ap11}
\hat X_k^{t+1}\leftarrow S_{\delta_{1k}^*,\delta_{2k}^*}\left(\hat X_k^t+\displaystyle\frac{\left< \hat r^t, \phi_k\right>}{\left< \phi_k, \phi_k\right>} \right)
\end{equation}
where $\hat X_k^{t}$ denote the estimator of $X_k$ in the $t$th step, and $\hat r^t=Y-\Phi \hat X^t$. The overall algorithm operates by applying this update repeatedly in a cyclical manner, updating the coordinates of $\hat X$ along the way. Once an initial estimator of $X$ is given, for example by lasso estimation or ridge estimation, the update can be continued until convergence.

\bibliography{biblio}
	
\end{document}